\documentclass[twocolumn,preprintnumbers,amsmath,amssymb,superscriptaddress]{revtex4}
\usepackage{graphicx}
\usepackage{dcolumn}
\usepackage{bm}
\usepackage{color}
\usepackage{subfigure}
\usepackage{amsbsy}

\usepackage{hyperref}
\hypersetup{
    colorlinks=true,
    linkcolor=blue,
    filecolor=blue,
    citecolor=blue,
    urlcolor=blue,
}

\begin{document}

\title{Fast electrical modulation of strong near-field interactions \\between erbium emitters and graphene}

\author{Daniel Cano}
\affiliation{ICFO - The Institute of Photonic Sciences, Castelldefels 08860, Barcelona, Spain}

\author{Alban Ferrier}
\affiliation{Institut de Recherche de Chimie Paris (IRCP), Universit\'{e} PSL, Chimie ParisTech, CNRS, 75005 Paris, France}
\affiliation{Facult\'{e} des Sciences et Ing\'{e}nierie, Sorbonne Universit\'{e}s, UFR 933, 75005 Paris, France}

\author{Karuppasamy Soundarapandian}
\affiliation{ICFO - The Institute of Photonic Sciences, Castelldefels 08860, Barcelona, Spain}
\author{Antoine Reserbat-Plantey}
\affiliation{ICFO - The Institute of Photonic Sciences, Castelldefels 08860, Barcelona, Spain}
\author{Marion Scarafagio}
\affiliation{Institut de Recherche de Chimie Paris (IRCP), Universit\'{e} PSL, Chimie ParisTech, CNRS, 75005 Paris, France}
\author{Alexandre Tallaire}
\affiliation{Institut de Recherche de Chimie Paris (IRCP), Universit\'{e} PSL, Chimie ParisTech, CNRS, 75005 Paris, France}
\author{Antoine Seyeux}
\affiliation{Institut de Recherche de Chimie Paris (IRCP), Universit\'{e} PSL, Chimie ParisTech, CNRS, 75005 Paris, France}
\author{Philippe Marcus}
\affiliation{Institut de Recherche de Chimie Paris (IRCP), Universit\'{e} PSL, Chimie ParisTech, CNRS, 75005 Paris, France}
\author{Hugues de Riedmatten}
\affiliation{ICFO - The Institute of Photonic Sciences, Castelldefels 08860, Barcelona, Spain}
\affiliation{ICREA - Instituci\'o Catalana de Re\c{c}erca i Estudis Avancats, 08010 Barcelona, Spain}
\author{Philippe Goldner}
\affiliation{Institut de Recherche de Chimie Paris (IRCP), Universit\'{e} PSL, Chimie ParisTech, CNRS, 75005 Paris, France}
\author{Frank H. L. Koppens}
\affiliation{ICFO - The Institute of Photonic Sciences, Castelldefels 08860, Barcelona, Spain}
\affiliation{ICREA - Instituci\'o Catalana de Re\c{c}erca i Estudis Avancats, 08010 Barcelona, Spain}
\author{Klaas-Jan Tielrooij}
\affiliation{Catalan Institute of Nanoscience and Nanotechnology (ICN2), BIST and CSIC, Campus UAB, Bellaterra (Barcelona), 08193, Spain}

\begin{abstract}
Combining the quantum optical properties of single-photon emitters with the strong near-field interactions available in nanophotonic and plasmonic systems is a powerful way of creating quantum manipulation and metrological functionalities. The ability to actively and dynamically modulate emitter-environment interactions is of particular interest in this regard. While thermal, mechanical and optical modulation have been demonstrated, electrical modulation has remained an outstanding challenge. Here we realize fast, all-electrical modulation of the near-field interactions between a nanolayer of erbium emitters and graphene, by in-situ tuning the Fermi energy of graphene. We demonstrate strong interactions with a $>$1,000-fold increased decay rate for $\sim$25\% of the emitters, and electrically modulate these interactions with frequencies up to 300 kHz –- orders of magnitude faster than the emitter’s radiative decay ($\sim$100 Hz). This constitutes an enabling platform for integrated quantum technologies, opening routes to quantum entanglement generation by collective plasmon emission or photon emission with controlled waveform.
\end{abstract}

\maketitle
\section{INTRODUCTION}

One of the major accomplishments of nanophotonics research is the accurate control of the near-field interactions between light emitters and their environment. These interactions typically lead to a modification of the decay rate and emission properties, with decay-enhancement factors $F_{\rm P}$ up to 1,000 for emitters in the near-field of metallic nano-antennas\cite{Aks:14}. Such systems, however, do not easily allow for active modulation of the emitter-environment interactions. Active and dynamic control of emitter-environment interactions has been achieved in photonic crystal cavities and nano-electromechanical systems through modulation of the effective refractive index of the environment, using electromechanical actuation\cite{Cot:18}, mechanical oscillations\cite{Pla:16, Tia:16}, or photoexcitation of free carriers\cite{Jin:14}. These techniques, however, either only work with one fixed modulation frequency or require optical fields for fast modulation. Furthermore, the achieved decay-enhancements are typically moderate, with $F_{\rm P} \sim$10.

An effective solution can be found in graphene, a two-dimensional material that combines a high degree of electro-optical tunability at high speeds (tens of GHz\cite{Pha:15}) with strong near-field light-matter interactions. Graphene not only provides a wide range of electrical tunability through its Fermi energy. At sufficiently high Fermi energy it supports plasmons with much stronger field confinement than noble metals:
the wavelength of graphene plasmons $\lambda_{\rm pl}$ is $\sim$100 times smaller than that of free-space photons $\lambda_0$, whereas in metallic thin films $\lambda_0/\lambda_{\rm pl}$ is typically lower than 3 \cite{Nov:06}. For pristine graphene, this leads to a mode volume confinement of $(\lambda_0 / \lambda_{\rm pl})^3 \simeq 10^6$ \cite{Jab:09}. Moreover, it is possible to  reach a mode volume confinement up to $10^9$ in graphene-metal grating structures\cite{Alc:18} and $10^{10}$ in cube-grating structures\cite{Eps:20}, with potential for very strong near-field interactions with nearby emitters. Many suggested plasmon-based technologies, where the fields of photonics and electronics merge\cite{Mai:01, Gra:10}, would benefit from fast temporal control of emitter-plasmon interaction, including applications such as single-photon nano-antennas\cite{Koe:17}, and sub-diffraction limited sensors\cite{Bro:12}. Experimentally, however, active electrical tunability of emitter-graphene interaction has only been demonstrated with low decay-enhancement $(F_{\rm P} <$ 3) and in a non-dynamic fashion\cite{Lee:14, Tie:15}.

Here we present a hybrid system made of a nanoscale layer of erbium emitters in an oxide matrix, and graphene, in which the near-field interactions can be efficiently controlled on-chip, at high modulation frequencies, and by means of moderate electrical gate voltages (on the order of a few volt). Rare-earth erbium ions are technologically highly relevant as they emit photons at 1.54-$\mu$m wavelength, within the C-band of optical communication systems. Furthermore, rare-earth oxide crystals have a proven relevance in photonic quantum memories\cite{Bus:14} and spin-photon interfaces\cite{Kut:19}, while nanoscale rare-earth materials are currently gaining interest\cite{Zho:19}. We will show highly efficient erbium-graphene interaction with decay-enhancement factors $F_{\rm P}$ larger than 1,000 for $\sim$25\% of the ions, which means that $>$99.9\% of the energy of these excited ions flows to graphene through near-field interactions. Importantly, we demonstrate modulation of the erbium-graphene interaction on a much shorter time scale than the lifetime of the single-photon emitters. In this special dynamical case, the quantum regime emerges, with exciting possibilities such as generation of single photons with controlled waveform\cite{Thy:13, Kel:04}, Dicke phases\cite{Zha:19a,Zha:19b} and quantum entanglement generation by collective plasmon emission\cite{Man:12}.
\vspace{-2mm}
\section{RESULTS}
\vspace{-1mm}
\subsection{Hybrid erbium-graphene system with dual-gate modulation}
\vspace{-2mm}
We show the design of our hybrid system schematically in Fig.\ 1. The central part consists of a graphene monolayer on a nanoscale layer of erbium-doped Y$_2$O$_3$ (2\%). To achieve strong near-field interactions, the erbium emitters should be located at nanoscale distances from the 2D material -- ideally within the sub-wavelength volume occupied by the highly-confined plasmons ($<$15 nm). A sufficiently thin layer with erbium emitters is therefore required. However, crystals of nanoscale dimensions typically suffer from detrimental non-radiative losses due to surface defects\cite{Dor:12}. Removing this loss mechanism is crucial, as it constitutes a competing energy flow channel for erbium-graphene interactions that would hinder the observation of these interactions. We overcome this experimental bottleneck by using atomic layer deposition (ALD)\cite{Sca:19} with optimized post-treatment -- a technique that produces few-nanometer-thick rare-earth-doped Y$_2$O$_3$ films with atomic scale thickness control and emission quality as in bulk crystals (see Supplementary Note 1, Supplementary Figs.\ 1-2 and Supplementary Table 1). The results we will show are obtained with a sample containing an erbium layer of 12 nm thickness, as measured through white light interferometry (see Supplementary Note 2 and Supplementary Figs.\ 3-4 for more sample characterization).


With the aim of dynamically controlling the erbium-graphene interactions through the Fermi energy of graphene $E_{\rm F}$, we combine a backgate of $p$-doped silicon with a polymer electrolyte topgate \cite{Das:08} (see Fig.\ 1). The backgate, with a smaller AC impedance than the topgate, is very suitable for high-frequency modulation. We use it to modulate $E_{\rm F}$ at high frequencies over a range of $\sim$0.3 eV. On the other hand, the topgate allows Fermi energy tuning over a range larger than 1 eV, which is sufficiently high to launch plasmons in resonance with the photons emitted by erbium, whose energy is $E_{\rm Er} = 0.8$ eV (see Fig.\ 1a). We use the topgate to provide a high base Fermi energy during high-frequency modulation of the backgate. The Fermi energies induced by both gates are calibrated by Hall measurements and resistance measurements (see Supplementary Note 3 and Supplementary Fig.\ 5). Thus, using our dual-gated, hybrid erbium-graphene system we can modulate $E_{\rm F}$, for example, between 0.6 eV and 0.3 eV, (Fig.\ 1b). This modulates the system between two regimes, where the erbium emitters decay by transferring energy to graphene, leading to intraband absorption (Fig.\ 1a) and interband absorption (Fig.\ 1c), respectively. The intraband regime is mainly associated with plasmon generation in graphene, whereas in the interband regime mainly electron-hole pair creation occurs.
\vspace{-1mm}
\subsection{Emission contrast and decay enhancement}
\vspace{-2mm}
The intraband and interband absorption regimes can be experimentally distinguished because they are associated with different local densities of optical states (LDOS), leading to distinct decay rates for the emitters innteracting with graphene. Our calculations following Refs.\cite{Kop:11,Gon:16} show that the decay-enhancement factor $F_{\rm P}$ for an emitter at 5 nm from graphene is $\sim$100 in the intraband regime ($E_{\rm F}$ = 0.6 eV) and $>$1,000 in the interband regime ($E_{\rm F}$ = 0.3 eV). In order to observe these regimes experimentally, we excite the erbium ions with a 532-nm laser and detect their emission at 1.54 $\mu$m in a scanning confocal microscope (see Methods), while varying the Fermi energy using the topgate. Figure 2a shows the measured emission contrast, defined as the emission without graphene (measured by shining the laser outside the graphene channel) divided by the emission with graphene (measured by shining the laser on graphene; see also Supplementary Fig.\ 3). The excitation power is sufficiently low (fluence of $10^4$ W cm$^{-2}$) to ensure that the ion transition is not saturated (see Supplementary Fig.\ 4). We indeed identify two different regimes: At low Fermi energies ($E_{\rm F} <$ 0.4 eV), energy transfer is mostly caused by interband transitions in graphene. Above 0.4 eV, interband transitions become suppressed by Pauli blocking, because $E_{\rm F}>E_{\rm Er}/2$, and intraband transitions become the dominant energy decay pathway of the ions. The positive slope of the emission contrast for $E_{\rm F}>$ 0.6 eV is a clear signature of the presence of plasmons. The slope is positive because the decay length of the plasmon field is approximately the same as the plasmon wavelength, $\lambda_{\rm pl}$, which scales linearly with $E_{\rm F}$. Therefore, as $E_{\rm F}$ increases, the volume occupied by the plasmon field increases, thus increasing the number of ions interacting with plasmons, and thereby decreasing the amount of emitted light.

To obtain more insight into the dynamics of the erbium-graphene interactions, we measure the erbium emission decay curves. We modulate the excitation laser into pulses and use single-photon counting electronics to create emission histograms (see Methods). Figure 2b shows the decay curves measured with the laser shining on graphene, for the interband transition regime ($E_{\rm F}\sim 0.2$ eV) and for the intraband regime ($E_{\rm F}\sim 0.8$ eV), as well as in a region without graphene. We observe faster decay in the intraband regime (e$^{-1}$ time of $\sim$3 ms) and even faster decay in the interband regime (e$^{-1}$ time of $\sim$1 ms), compared to the decay without graphene (e$^{-1}$ time of $\sim$6 ms), in qualitative agreement with the emission contrast measurements, which show less emission in the interband regime than in the intraband regime.

\onecolumngrid

\begin{figure}
\centerline{\scalebox{0.55}{\includegraphics{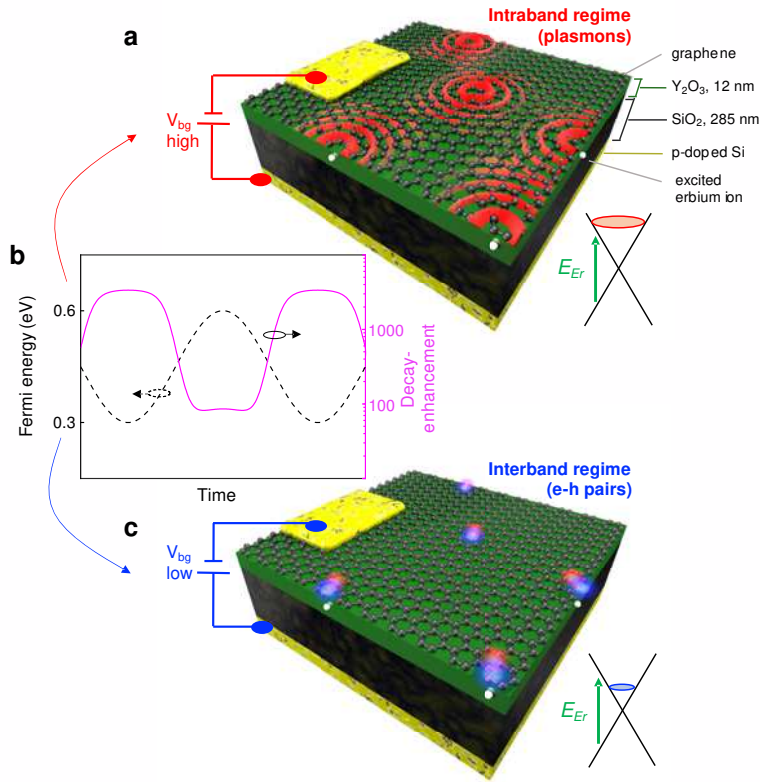}}}
\caption{\textbf{Concept of dynamic modulation of hybrid erbium-graphene system.}
\textbf{a)} Schematic illustration of the hybrid erbium-graphene system when the Fermi energy of graphene is tuned to $\sim$0.6 eV and the erbium-graphene interaction leads to intraband transitions, mainly associated with launching of propagating graphene plasmons (red waves). The system contains, from top to bottom, a monolayer of graphene, a thin film ($\sim$12 nm) of Y$_2$O$_3$ containing erbium ions (white spheres), a 285-nm-thick SiO$_2$ layer, and a $p$-doped silicon backgate. A backgate voltage, $V_{\rm bg}$, is applied between the backgate and a gold electrode in contact to graphene for fast modulation. The SiO$_2$ layer serves as electrical isolation between graphene and the p-doped silicon. On top of graphene, there is a transparent topgate made of polymer electrolyte (not shown in the image). \textbf{b)} Sinusoidal time evolution of the Fermi energy of graphene (dashed black line, left vertical axis) and the corresponding decay-enhancement factor $F_{\rm P}$ for an erbium emitter located at $z = 5$ nm from graphene (solid purple line, right vertical axis). The modulation of the Fermi energy, calculated following Refs.\ \cite{Kop:11, Gon:16}, leads to a modulation of $F_{\rm P}$ by more than one order of magnitude – from $<$100 in the intraband regime to $>$1,000 in the interband regime. \textbf{c)} Schematic illustration of the hybrid erbium-graphene system for $E_{\rm F}\sim$ 0.3 eV, which corresponds to interband transitions,  mainly creating electron-hole pairs (red-blue spheres).}
\end{figure}

\twocolumngrid

Strikingly, the decay curves are multi-exponential, with a large negative slope in the beginning of the decay (see inset of Fig.\ 2b). This multi-exponential behavior stems from the different decay rates $\gamma$ of the emitters, depending on their distance to graphene $z$, since $\gamma$ scales with $z^{-4}$ in the interband regime and with $\exp(- 4 \pi z/\lambda_{\rm pl})$ in the intraband regime\cite{Kop:11,Gau:13}. The ions located furthest away from graphene have the lowest energy transfer rates, thus emitting more photons and extending the decay curves to long times. The strongly negative slope in the beginning of the decay curves (with an estimated decay time $<$100 $\mu$s; see inset of Fig.\ 2b and Supplementary Fig.\ 10) is due to the high energy transfer rate for small $z$, and indicates the presence of a significant fraction of ions with very strong erbium-graphene interactions. Thus, the decay curves provide information of the $z$-dependence of the interactions, while the emission contrast measurements reflect the overall (distance-integrated) graphene-induced decay-enhancement $F_{\rm P}$, which is $\sim$7 ($\sim$2.5) in the interband (intraband) regime. This indicates that an overall fraction of $\eta \approx (F_{\rm P} - 1) / F_{\rm P} \approx$ 85\% (60\%) of the energy of excited erbium emitters flows to interband (intraband) transitions in graphene. We note that we have reproduced these decay curves with multiple erbium-graphene samples (see Supplementary Note 4 and Supplementary Fig.\ 9).

\null\newpage 

\onecolumngrid

\begin{figure}
\centerline{\scalebox{0.45}{\includegraphics{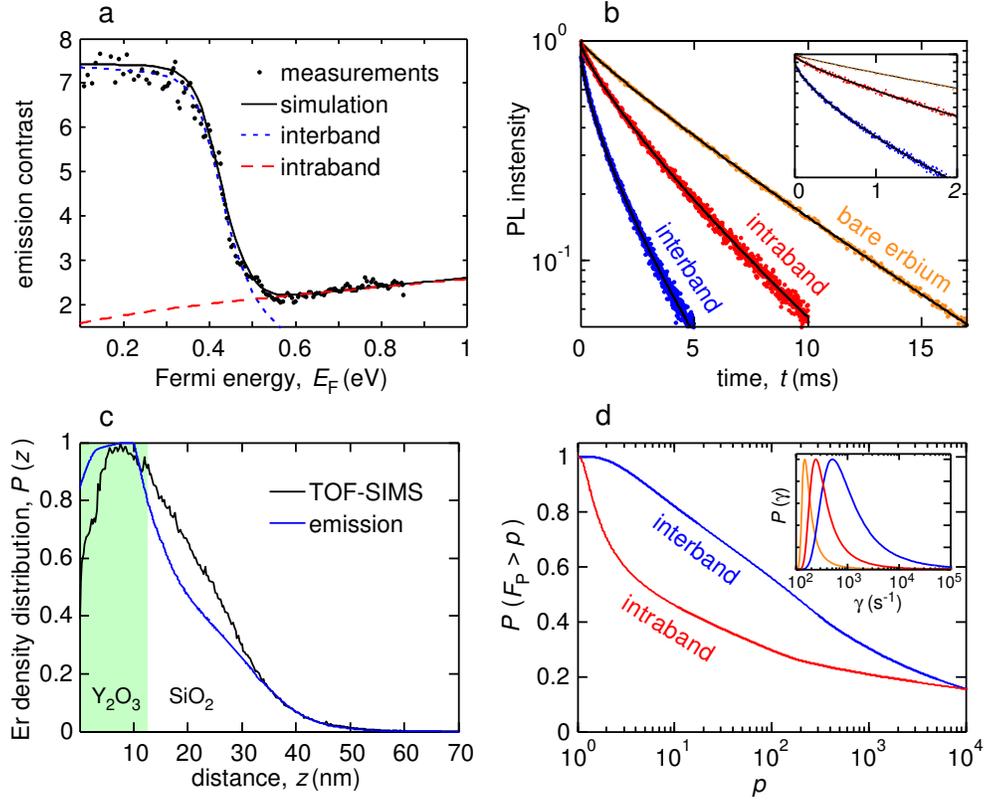}}}
\caption{\textbf{Efficient energy flow from erbium emitters to graphene.}
\textbf{a)} Measured emission contrast, defined as the emission with the excitation laser shining outside graphene, divided by the emission with the laser on graphene, as a function of the Fermi energy (black dots). The black solid line represents the calculated emission contrast from $N$ erbium ions located at different distances $z_i$ from graphene (see Methods), using the total graphene conductivity. The red (blue) dotted line represents the calculated emission contrast by only considering the conductivity of the intraband (interband) excitation, showing the different microscopic origins of the emitter-graphene interactions at high and low $E_{\rm F}$. The interband contribution (blue) quickly drops when $E_{\rm F} >$  0.4 eV, due to Pauli blocking, whereas the intraband contribution (red) steadily rises with $E_{\rm F}$.  \textbf{b)} Measured decay curves of erbium ions for three cases: graphene in the interband regime ($E_{\rm F}= 0.3$ eV, blue dots), graphene in the intraband regime ($E_{\rm F}= 0.8$ eV, red dots), and without graphene (orange dots). The black solid lines are the best-fit stretched-exponential functions. The inset focuses on the beginning of the decay curves, illustrating the multi-exponential behaviour. \textbf{c)} Erbium density distribution (normalized to maximum) \textit{vs.} erbium-graphene distance obtained from the combined analysis of the emission contrast and the decay curve of the interband regime (blue) and from TOF-SIMS measurements (black). \textbf{d)} Cumulative distribution of the decay enhancement factor calculated for the interband (blue) and the intraband (red) regimes. These curves show that about 80\% (50\%) of the ions have $F_{\rm P} > 10$, for the interband (intraband) regime, and approximately 25\% of ions have $F_{\rm P} >1,000$ for both regimes. The inset shows the total decay rate distributions, directly obtained from the multi-exponential fitting of the experimental decay curves (see Methods). The colors are the same as in panel \textbf{b}.}
\end{figure}

\twocolumngrid

\subsection{Quantifying erbium-graphene interaction}

Given the strong $z$-dependence of the emitter-graphene interaction, it is crucial to determine the distribution of erbium ions in order to quantitatively understand the energy transfer efficiency of the different erbium ions in the nanolayer. We analyze the emission contrast and the decay curves (Figs.\ 2a and b) together because they provide complementary information. The decay curves provide with high accuracy the distribution of ions with relatively low $F_{\rm P}$ factors (large $z$), as these are the ions that emit the largest number of photons during lifetime measurements. On the other hand, the emission contrast measurements of Fig.\ 2a reveal more accurately the effect of the ions with relatively large $F_{\rm P}$ (small $z$), as these are the ions that transfer the highest fraction of their energy to graphene, leading to the largest decrease in emission. Together, the decay curves and emission contrast measurements yield the density of erbium ions as a funcion of $z$ (see Fig.\ 2c).

In detail, we extract the distribution for ions with $z >$ 7 nm directly from the decay curves of Fig.\ 2b, by describing each multi-exponential decay curve by a continuous sum of exponentially decaying functions, whose probability amplitudes are given by the decay-rate distribution $P(\gamma)$. In terms of the ion dynamics, the distribution $P(\gamma)$ can be interpreted as the likelihood that a given ion decays with a certain decay rate $\gamma$. The inset of Fig.\ 2d clearly shows that the decay-rate distribution $P(\gamma)$ is shifted towards higher values of $\gamma$ in the regions of the device with graphene, indicating that the decay rate increases due to energy transfer to graphene. From the analysis of these decay-rate distributions, we  extract the distribution of decay-enhancement factors $P(F_{\rm P})$, following the numerical procedure described in the Methods. Then, we convert $P(F_{\rm P})$ into the distribution of erbium-graphene distances $P(z)$, i.e. the density distribution, by using the theoretical relation between $F_{\rm P}$ and the emitter-graphene distance $z$.

It turns out that the distributions $P(F_{\rm P})$ and $P(z)$ obtained from the decay curves are accurate up to $F_{\rm P} = 10^3$, corresponding to $z \gtrsim $ 7 nm. For higher values of $F_{\rm P}$, the decay is so fast and has such a small amplitude, that we cannot resolve it experimentally in our decay curves (see Supplementary Note 5). However, we can obtain the distribution of the ions with $z < $ 7 nm using the emission contrast measurements of Fig.\ 2a. For this, we use a computational model of $N$ ions at different distances from graphene, $z_i$ ($i$ = 1,...,$N$), and find the ion density distribution $P(z)$ that best reproduces the measurements of Figs.\ 2a and b (see Methods). This is how we obtain the ion density distribution in Fig.\ 2c.

We compare the ion distribution extracted from optical measurements $P(z)$ with the density distribution measured by means of time-of-flight secondary ion mass spectrometry (see Methods and Supplementary Note 6). The similarity between the two density distributions confirms the validity of our analysis. Interestingly, our results indicate that some ions have diffused from the Y$_2$O$_3$ layer into the underlying SiO$_2$ layer. This has likely occurred during the annealing post-treatment of the films. These diffused ions interact less strongly with graphene and lead to the moderate overall emission contrast we observe in Fig.\ 2a.

Importantly, our analysis of the experimental data of Figs.\ 2a-b provides evidence of very strong emitter-graphene interactions at the shortest distances. Figure 2d shows the calculated cumulative distribution of $P(F_{\rm P})$,
\begin{equation}
P(F_{\rm P} > p) = \int_p^{\infty} P(F_{\rm P}) {\rm d} F_{\rm P},
\label{Eq_cumulative distribution}
\end{equation}
which describes the probability that an ion has a decay-enhancement factor larger than $p$. In this way, the cumulative distribution $P(F_{\rm P} > p)$ provides the fraction of ions with $F_{\rm P} > p$. Using Equation (1), we find that about 80\% (50\%) of the ions have decay enhancement factor $F_{\rm P} >$ 10 for the interband  (intraband) regime, and approximately 25\% of ions have $F_{\rm P} >$ 1,000 for both regimes, which means that more than 99.9\% of the energy from these ions flows to graphene.

\subsection{Fast electrical modulation of near-field interactions}

Having established the occurrence of highly efficient erbium-graphene interactions in our system, we now demonstrate dynamic control of these interactions on a time scale that is much shorter than the emitter's lifetime  of $\sim$10 ms \cite{Web:68}. We induce a fast temporal variation in the LDOS experienced by the emitters by modulating the Fermi energy of graphene. To this end, we apply an AC voltage to the backgate, and verify the effect of this modulation on the excited state populations of erbium by measuring the temporal variations of photon emission using a single-photon counting setup. In these experiments, we keep the excitation laser power constant. In Fig.\ 3, we modulate the Fermi energy between 0.3 eV and 0.6 eV at different modulation frequencies $f_{\rm mod}$ between 20 Hz and 300 kHz. This corresponds to a modulation of the erbium decay pathway between interband and intraband transitions, as in Fig.\ 1. In these measurements, we thus establish control over the timing of plasmon launching from erbium ions down to the microsecond range, which is remarkable for emitters with millisecond natural lifetime. We verified that there is no dynamic response outside graphene and that a backgate voltage of $<$10 V is sufficient for complete modulation between the two emitter-graphene interaction regimes (see Supplementary Fig.\ 6).

As $f_{\rm mod}$ increases and becomes higher than the emitter decay rate, the internal dynamics of the ions are not able to follow the temporal variations of the environment. This results in a gradual delay of the maximum and minimum of the time-dependent emission with respect to the sinusoidal oscillation of the Fermi energy and the reduction of the emission modulation amplitude, which depends on $1/f_{\rm mod}$ whenever $f_{\rm mod} \gg \gamma$. Interestingly, these modulation frequencies surpass not only the decay rate of the ions, but also the quantum coherence decay rate of erbium in Y$_2$O$_3$ (11 kHz; measured at 2.5 K in a bulk ceramic sample and under a small external magnetic field of 0.65 T)\cite{Zha:17}.

In a second series of measurements, shown in Fig.\ 4, we modulate the Fermi energy between 0.7 eV and 0.9 eV. Here we apply a higher voltage to the polymer electrolyte topgate than in the previous modulation experiment, while modulating the backgate again with a voltage of $<$10 V. In this situation, the system is always in the intraband regime, where plasmon launching is the dominant energy decay pathway, and we therefore modulate the strength of the emitter-plasmon interaction. Note that an increase in backgate voltage now leads to a decrease in emission, because stronger emitter-plasmon interaction leads to less emission (see also Supplementary Fig.\ 7). This is in contrast with the case in Fig.\ 3, where an increase in backgate voltage leads to a transition from the interband absorption regime to the intraband regime, giving more emission.

\begin{figure}
\centerline{\scalebox{0.35}{\includegraphics{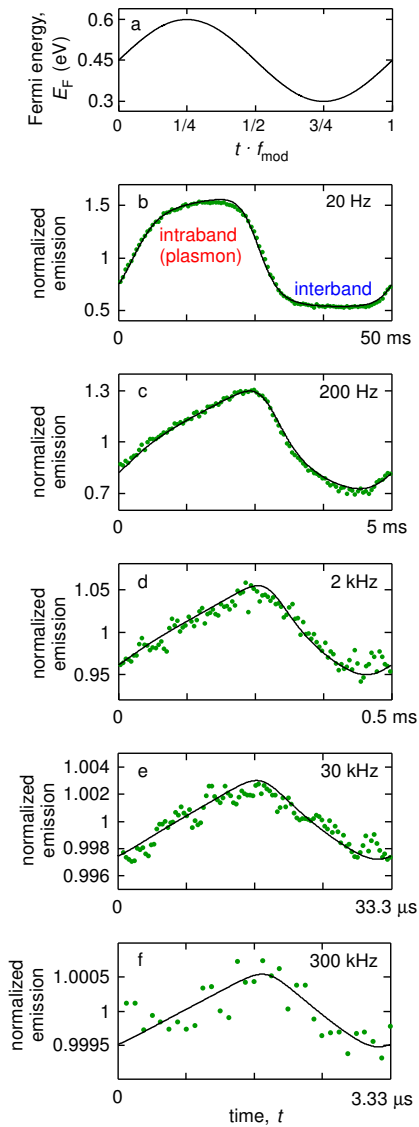}}}
\caption{\textbf{Dynamic modulation between interband and intraband regimes.}
\textbf{(a)} Fermi energy as a function of time $t$ using a sinusoidal function of frequency $f_{\rm mod}$ (schematic). The slow topgate is used to tune to $E_{\rm F}\sim 0.45$ eV, whereas the fast backgate provides the modulation between 0.3 eV and 0.6 eV. \textbf{b-f)} Time-resolved photon emission while the Fermi energy is modulated at different frequencies between 20 Hz and 300 kHz (green dots). Every time-modulated emission measurement is normalized to its mean value, so that the modulation does not depend on the excitation laser power or the photon collection efficiency. The black solid curves show the dynamic simulations of $N$ ions located at the distribution of distances from graphene $P(z)$, obtained from the emission contrast measurements and decay curves (see Methods). Note that panel \textbf{f)} shows modulation on a timescale of microseconds, whereas the radiative lifetime of erbium ions in Y$_2$O$_3$ is $\sim$10 ms, a difference of 4 orders of magnitude.}
\end{figure}

\begin{figure}
\centerline{\scalebox{0.35}{\includegraphics{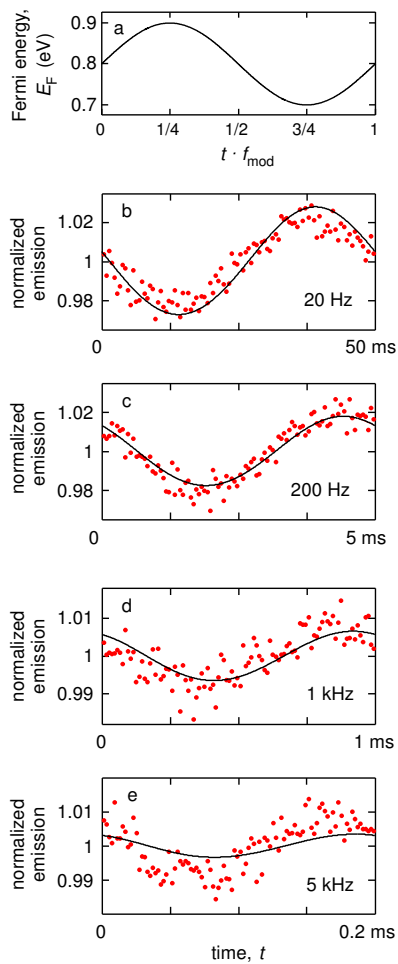}}}
\caption{\textbf{Dynamic modulation within the intraband regime.}
\textbf{a)} Fermi energy as a function of time $t$ using a sinusoidal function of frequency $f_{\rm mod}$ (schematic). The slow topgate is used to tune to $E_{\rm F}\sim 0.8$ eV, whereas the fast backgate provides the modulation between 0.7 eV and 0.9 eV, thus modulating the interaction strength within the intraband regime. \textbf{b-e)} Time-resolved photon emission while the Fermi energy is modulated at different frequencies between 20 Hz and 5 kHz (red dots). Every time-modulated emission measurement is normalized to its mean value, so that the modulation does not depend on the excitation laser power or the photon collection efficiency. The black solid curves show the dynamic simulations of $N$ ions located at the distribution of distances from graphene $P(z)$ obtained from the emission contrast measurements and decay curves (see Methods). Note that the modulated emission is “out-of-phase” with the modulating Fermi energy, because a higher Fermi energy gives lower emission in the intraband regime. }
\end{figure}

\null\newpage 

We model the temporal modulation of photon emission by simulating the dynamics of $N$ ions using the distribution of ion distances $P(z)$ obtained from the emission contrast measurements and decay curves. For every ion $i$ at distance $z_i$, we numerically solve the rate equation with the time-dependent $F_{\rm P}$ factor and for the corresponding Fermi energy modulation (see Methods). We find good agreement between experimental data and numerical simulations, which do not contain any freely adjustable fit parameters (see Methods), adding credibility to our computational approach. We note that only in the case of Fig.\ 4e, we observe a larger modulation amplitude in the experiment than in the simulation. We speculate that this could be related to additional (e.g.\ non-local) effects in the plasmonic local field at very short distances from graphene.

\section{DISCUSSION}

We have demonstrated a material platform that integrates the optical properties of erbium quantum emitters with the strong near-field interactions of graphene, providing the conditions for quantum manipulation and metrological functionalities \cite{Sch:08,Kur:15}. In particular, our hybrid erbium-graphene platform enables fast temporal control over the strong near-field interactions, thus providing an efficient way to manipulate quantum states in nanoscale solid-state devices by means of conventional electronics. The control over the dynamics of single-photon emitters on a much shorter time scale than their lifetime is an essential ingredient towards observing intriguing effects such as Dicke phase transitions \cite{Zha:19a, Zha:19b}, non-linear light-matter interactions at the quantum level \cite{Cox:18}, and multi-particle entanglement generation \cite{Man:12}. In addition, such a fast control over the near-field interactions will expand existing functionalities of plasmonic nanodevices integrating graphene waveguides and cavities \cite{Kop:11,Gon:16,Tam:13,Alo:19,Chr:12}. It enables the control of the waveform of photons and plasmons emitted into a guided mode\cite{Morin:19,Kel:04}, which is a required capability for distributed quantum systems. It furthermore facilitates the connection of optical transistors\cite{Hwa:09,Gei:13} based on quantum emitters in integrated plasmonic circuits, with the possibility to attenuate or amplify the plasmon emission by means of a gate voltage. These promising applications will be boosted by the prospects of coupling plasmonic modes to the far field through optical nanoantennas or through optical waveguides \cite{Tiecke:15}, thus enabling controlled photon emission enhancement for optical communications. Overall, these prospects will stimulate the development of emitter-graphene interfaces as a building block for hybrid systems with applications in optoelectronic quantum technologies.

\vspace{5mm}

\section*{METHODS}

\subsection{Fabrication of erbium-graphene hybrid devices.} The erbium-yttria thin film depositions are carried out by ALD with a commercial reactor using conventional $\beta$-diketonate precursors: Er(tmhd)3 and Y(tmhd)3. The precursors are held at 160 $^\circ$C and delivered using N$_2$ as a carrier gas and O$_3$ as an oxidizing agent. The precursors are flown sequentially with 3 s injection time into the thermalized deposition chamber at 350 $^\circ$C. The number of cycles is adjusted in order to obtain the desired thickness (circa 12 nm here). The films are annealed at high temperature (950 $^\circ$C) for 2h in air prior to measurement in order to improve crystallinity. More details about the optimization of the growth procedure are discussed in  Supplementary Note 1. We use a moderate erbium concentration of 2\% in order to avoid possible non-radiative decay channels induced by erbium-erbium interactions. Single-layer graphene, grown by chemical vapour deposition (CVD) on copper, is transferred directly onto the surface of the 12-nm-thick Y$_2$O$_3$:Er$^{3+}$ layer using the standard wet transfer method. Graphene is patterned into a Hall-bar geometry by laser writing optical lithography and subsequent reactive ion etching (see Supplementary Fig.\ 3). The whole surface is covered with a transparent polymer electrolyte that serves as topgate. The electrolyte is made with polyethylene oxide (PEO) and LiClO$_4$ with 8:1 weight ratio in a solution of methanol\cite{Das:08}. The device contains six electrical contacts to graphene and two electrical contacts to apply the topgate voltage $V_{\rm tg}$ to the polymer electrolyte. The electrical contacts consist of a 50-nm-thick gold layer deposited on a 5-nm-thick chromium sticking layer and patterned by laser writing optical lithography. There is a 50-nm-thick SiO$_2$ protection layer between the gold electrodes and the electrolyte in order to isolate the gold electrodes from the polymer electrolyte.

\subsection{Experimental setup for optical measurements.} The experiments are carried out in a home-built scanning confocal microscope setup, with the sample mounted inside a vacuum chamber at a pressure of 5-10 mbar for optimal operation of the polymer electrolyte gate. The optical setup uses an infrared objective (Olympus LCPLN-50X-IR, numerical aperture 0.65), which provides a spatial resolution of $\sim$1 $\mu$m. A focused 532-nm laser beam, with typically a power of $\sim$0.2 mW at the sample, excites the ions into the short-lived state $^2$H$_{11/2}$, from which the erbium population rapidly decays into the metastable first excited state $^4$I$_{13/2}$ via nonradiative multiphonon emission. The emission of the $^4$I$_{13/2} \rightarrow ^4$I$_{15/2}$ transition at the characteristic wavelength 1.54 $\mu$m is collected, spectrally filtered with a narrow bandpass filter (Thorlabs FB1550-40) and directed into a near-infrared single-photon detector (ID Quantique id210) with very low level of dark counts ($\sim 9$ Hz). For time-resolved measurements, emission histograms are obtained using photon-counting electronics (PicoHarp 300) in time-tagged time-resolved acquisition mode, recording the arrival times of all photons. During lifetime measurements, the excitation laser intensity is modulated into square pulses by switching on and off the signal of an acousto-optic modulator. All measurements were carried out at room temperature. We verified that during measurements that can take up to several hours, the Fermi energy did not vary significantly (see Supplementary Fig.\ 8).

\subsection{Decay rate distributions from the decay curves.} The decay rate in the regions of the device without graphene is $\gamma_{\rm Er}=\gamma_{\rm ed}+\gamma_{\rm md}+\gamma_{\rm nr}$, where $\gamma_{\rm ed}\sim$75 Hz and $\gamma_{\rm md} \sim$50 Hz correspond to the electric and magnetic dipole moments, respectively\cite{Web:68}, and $\gamma_{\rm nr}$ represents the non-radiative decay channels of the erbium thin film. In the regions with graphene, the total decay rate is $\gamma=\gamma_{\rm Er}+\gamma_{\rm gr}$, where $\gamma_{\rm gr}$ is the rate of energy transfer to graphene. Every experimental decay curve, $n(t)$, is described as a continuous sum of exponential decays, $n(t)=\int_{0}^{\infty} \frac{P(\gamma)}{\gamma} e^{-\gamma t} d\gamma$, where $P(\gamma)$ is the probability distribution that describes the likelihood that a given ion decays at a certain rate $\gamma$. Here, we have considered that the excited erbium population, and thus the emission, is inversely proportional to the decay rate. The integral has the form of a Laplace transformation, so we can extract $P(\gamma)$
by inverse Laplace transformation using the numerical techniques of Ref.\ \cite{Joh:06} (see the inset of Fig.\ 2d). Next, we calculate the energy transfer rate distribution $P_{\rm gr}(\gamma_{\rm gr})$ by numerically solving the convolution equation, $P(\gamma)=\int_{0}^{\infty} P_{\rm Er}(\gamma-\gamma_{\rm gr})P_{\rm gr}(\gamma_{\rm gr}) d\gamma_{\rm gr}$. Here, $P_{\rm Er}(\gamma_{\rm Er})$ denotes the decay-rate distribution in the regions of the device without graphene (we write the subindex Er to indicate that the distribution only includes the intrinsic decay mechanisms of the erbium film). In doing the deconvolution, we filter out the small effect of the undesired non-radiative decay channels $\gamma_{\rm nr}$, thus obtaining the pure contribution of the erbium-graphene interactions, $\gamma_{\rm gr}$. The distribution $P_{\rm gr}(\gamma_{\rm gr})$ can be easily converted into the distribution of decay-enhancement factors, $P(F_{\rm P})$, by using the relation $\gamma_{\rm gr} = (F_{\rm P} -1) \gamma_{\rm ed}$. Next, we translate $P(F_{\rm P})$ into the density distribution, $P(z) = P(F_{\rm P}) $ d$F_{\rm P}/$d$z$. For this, the decay-enhancement factor as a function of distance, $F_{\rm P}(z)$, is calculated as in Refs.\ \cite{Kop:11,Gon:16}, where the response of graphene to the localized field of the emitter is simulated using the optical conductivity of the Kubo model, with a typical momentum scattering time of $\tau_{\rm sc}$ = 50 fs, and a refractive index of 1.8 for Y$_2$O$_3$ and 1.4 for the electrolyte (see Supplementary Fig.\ 9). The distributions obtained from the decay curves are very accurate for the lowest decay rate enhancements factors, $F_{\rm P} \lesssim 1,000$ . This cutoff is determined from the accuracy of the numerical inverse Laplace transform (see Supplementary Note 5). Ions with larger decay rates emit so few photons that they barely affect the slope of the decay curves, and their energy-transfer rates have to be investigated by using the emission contrast measurements. We have repeated the whole procedure using the decay curves of different Fermi energies (0.3 eV and 0.8 eV), and the results are practically the same.

\subsection{Decay rate distributions using the $N$-ion model.} We use a computational model of $N$ ions located at different distances from graphene, $z_i$ ($i = 1,...,N$). The positions $z_i \gtrsim  7$~nm are obtained by discretization of the density distribution $P(z)$ calculated from the decay curves. The positions $z_i \lesssim  7$~nm, are free parameters that we vary to find the distribution of distances, $\{z_i \}_{i=1,…,N}$, that best reproduces the emission contrast measurements of Fig.\ 2a. This variational procedure assumes that the density distribution is smooth. The calculations are accomplished by considering that the emission from every ion is proportional to its excited-state population, which in turn is proportional to $1/(F_{\rm P} \gamma_{\rm ed}+\gamma_{\rm md}+\gamma_{\rm nr})$, where the theoretical decay-enhancement factor $F_{\rm P} (z_i)$ is computed from the methods of Refs.\ \cite{Kop:11, Gon:16} (see above). Here, we assume that $\gamma_{\rm nr}=10$ Hz for all ions (see Supplementary Note 1). The discrete distribution $\{z_i \}_{i=1,…,N}$ is converted into a continuous distribution, $P(z)$, and vice versa, by integration and discretization, respectively. We typically use $N$ = 50, where the density distribution already nicely converges.

\subsection{Simulation of the dynamic response to gate modulation.} We simulate the quantum-state population dynamics for every ion $i$ of our $N$-ion model by numerically solving the rate equations \cite{Foo:05},

\begin{equation}
\frac{d N_e^{(i)}}{dt} = - N_{\rm e}^{(i)} \left[ F_{\rm P}^{(i)} \gamma_{\rm ed} + \gamma_{\rm md} + \gamma_{\rm nr} \right] + N_{\rm g}^{(i)} \gamma_{\rm exc},
\end{equation}
\begin{equation}
N_{\rm g}^{(i)} +  N_{\rm e}^{(i)} = 1,
\end{equation}
\noindent
where $N_{\rm g}^{(i)}$ and $N_{\rm e}^{(i)}$ are respectively the populations in ground and excited states, $F_{\rm P}^{(i)}$ is the time-dependent decay-enhancement factor at the position $z_i$, and $\gamma_{\rm exc}$ is the excitation rate, which depends on the excitation laser power. The results of the simulations do not depend on $\gamma_{\rm exc}$ since every measurement is normalized to its mean value, although we need to assume a certain value of $\gamma_{\rm exc}$ to do the computation. To calculate $F_{\rm P}^{(i)}$, we first obtain the oscillating energy $E_{\rm F}$ that corresponds to the applied gate voltages using the calibration described in Supplementary Note 3. The Fermi energy amplitude as a function of the back-gate voltage amplitude is $\Delta E_{\rm F}= B \Delta V_{\rm bg}$, where $B = 15$ eV mV$^{-1}$ (13 eV mV$^{-1}$) in the experiments of Fig.\ 3 (Fig.\ 4). We then convert $E_{\rm F}$into $F_{\rm P}$ using the methods of Refs.\ \cite{Kop:11, Gon:16} (see above).

\subsection{Time-of-Flight Secondary Ion Mass Spectrometry (ToF-SIMS).} We use a dual-beam ToF-SIMS spectrometer (IONTOF GmbH, M\H{u}nster, Germany) to measure the density profile of the yttrium and erbium ions at a sufficiently low primary ion dose density to keep static conditions. The spectrometer was operated at a pressure of 10$^{-9}$ mbar. A pulsed 25 kV Bi$^+$ primary ion beam delivering 1 pA over a 100$\times$100 $\mu$m$^2$ area is used to etch the chemical species from the surface. The masses of the removed chemical species are determined by time-of-flight mass spectroscopy. The sputtering of the surface was done using a 2 keV Cs$^+$ sputter gun giving a 100 nA target current over a 300$\times$300 $\mu$m$^2$ area. The interlacing between Bi$^+$ and Cs$^+$ guns allows to record TOF-SIMS depth profiles. We used the profile of the removed YO$^-$ particles to extract the Er$^{3+}$ density distribution. This approximation is justified since the diffusion coefficients of yttrium and erbium are practically the same (see Supplementary Fig.\ 11).

\section*{ACKNOWLEDGEMENTS}

We thank Javier Garc\'{i}a de Abajo and David Hunger for discussions, and Juan Sierra, Marius Costache and Sergio Valenzuela for assistance with the Hall measurements. K.J.T. acknowledges funding from the European Union’s Horizon 2020 research and innovation programme under Grant Agreement No. 804349. ICN2 was supported by the Severo Ochoa program from Spanish MINECO (Grant No. SEV-2017-0706). This project has received funding from the European Union’s Horizon 2020 research and innovation programme under grant agreement No 712721 (NanOQTech).



\end{document}


\onecolumngrid

\noindent \textbf{\normalsize{SUPPLEMENTARY MATERIAL: FAST ELECTRICAL MODULATION OF STRONG NEAR-FIELD INTERACTIONS BETWEEN ERBIUM EMITTERS AND GRAPHENE}}
\vspace{1cm}
\twocolumngrid

\setcounter{page}{11}

\section*{Supplementary Note 1. Emission properties of erbium-doped thin films}

Removing the intrinsic non-radiative decay from the erbium-doped Y$_2$O$_3$ films is an essential requirement for the accurate experimental evaluation of the energy transfer at the erbium-graphene interface. The non-radiative losses due to defects that are usually present in nanoscale rare-earth-doped crystals constitute a competing energy flow channel for the erbium-graphene interactions, leading to undesired emission quenching. We overcome this problem by using a fabrication procedure that we have optimized to produce few-nanometer-thick erbium-doped Y$_2$O$_3$ films with almost pure radiative decay. In this fabrication procedure, the erbium-doped Y$_2$O$_3$ films are grown by atomic layer deposition (ALD), see Scarafagio \textit{et al.}\cite{Sca:19} for details, with an optimized annealing post-treatment that eliminates the non-radiative decay channels. With this technique, we are able to exploit the many advantages of ALD, such as accurate thickness control at atomic scale, good uniformity and the ease to vary the erbium doping level, while preserving the high optical quality of the erbium emitters.

\begin{figure}
\centerline{\scalebox{0.35}{\includegraphics{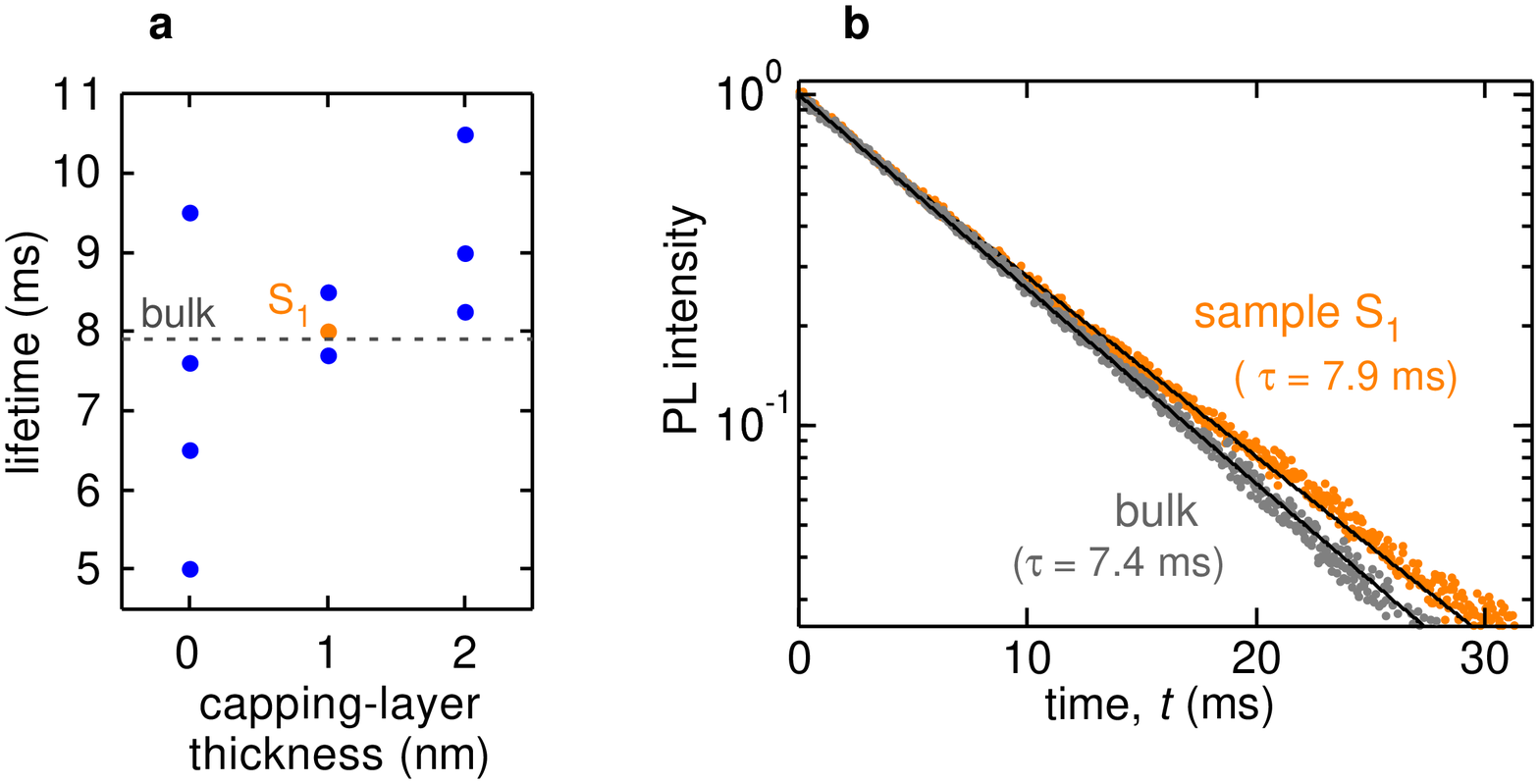}}}
\caption{\textbf{Comparison between thin-film samples and the bulk reference sample.} \textbf{(a)} Lifetimes of ten different thin film samples consisting of a 11-nm-thick Y$_2$O$_3$:Er (2\%) film grown by ALD on a SiO$_2$/Si wafer, with subsequent annealing at 950~$^\circ$C for 2 hours. Six thin film samples had an additional Y$_2$O$_3$ capping layer, annealed at 600~$^\circ$C for 2 hours, the thickness of which was 1 nm in three samples and 2 nm in the other three. These measurements suggest that the capping layer reduces the probability of non-radiative decay channels. The dashed line shows the lifetime of the reference bulk sample. \textbf{(b)} Decay curves of the thin film sample S$_1$ (before depositing graphene) and of bulk erbium-doped Y$_2$O$_3$. The experimental data is described by a mono-exponential curve of lifetime $\tau$ (black solid line).}
\label{fig:S1_bulk}
\end{figure}

For this work, we first did preliminary tests with erbium-doped Y$_2$O$_3$ films on Si(100), for different thicknesses and annealing parameters, and determined that optimal annealing occurs at 950~$^\circ$C for 2 hours. Using these parameters, we produced a series of ten thin film samples made of a 11-nm-thick Y$_2$O$_3$:Er (2\%) film grown on a 285-nm-thick SiO$_2$ layer of a p-type silicon wafer. This oxide layer serves as electrical isolation between the graphene devices and the backgate. Afterwards, on six of the ten thin film samples we grew an additional Y$_2$O$_3$ capping layer with a thickness of either 1 nm or 2 nm, and we annealed the samples again at 600~$^\circ$C for 2 hours. Supplementary Figure \ref{fig:S1_bulk}a shows the measured lifetimes of the ten thin film samples, where we can see the beneficial impact of the capping layer in reducing the non-radiative decay channels. In order to verify that the non-radiative processes due to surface defects are negligible, we compared the decay curves of the thin film samples with the decay curve of a bulk reference sample of erbium-doped Y$_2$O$_3$. Except in two thin film samples without capping layer, the decay curves were almost mono-exponential and had similar, or even longer, lifetimes as compared to the bulk reference. As a bulk reference we used a ceramic prepared by high-temperature solid-state reaction. For this, high purity Y$_2$O$_3$ and Er$_2$O$_3$ were mixed and annealed twice with intermediate crushing at 1500~$^\circ$C for 24 hours.

For the main study, we chose a thin film sample with similar emission properties as the bulk reference material. Supplementary Figure \ref{fig:S1_bulk}b shows the decay curve of the thin film sample, which we call S$_1$, together with the decay curve of the bulk reference material. The similarity between the two almost mono-exponential decays is an evidence of the negligible effect of the surface defects in sample S$_1$. The slightly longer lifetime in the thin film sample is the result of the lower effective refractive index of the thin film environment in comparison with bulk Y$_2$O$_3$. The significant improvements of the thermal post-treatment are shown in Supplementary Fig. \ref{fig:S1_SA_SB_SC} and Supplementary Table 1, which compare the decay-rate probabilities of sample S$_1$ with other thin film samples made without optimized thermal post-treatment. In S$_1$, the rate at which de emission decays to 1/e is $\gamma_{\rm 1/e} \sim$ 134 Hz, just a bit larger than the decay rate of the electric and magnetic dipole moments, $\gamma_{\rm ed}+\gamma_{\rm md} \sim 125$ Hz (Supplementary Reference 2). This shows that the decay is dominated by radiative processes. Furthermore, the full width half maximum (FWHM) of the decay rate distribution is only 33 Hz, much lower than $\gamma_{\rm 1/e}$. This is the evidence that the ion ensemble is highly homogeneous and that the exponential decay curves are hardly affected by detrimental effects such as non-radiative decay channels, erbium-erbium interactions and energy migration to quenching centers. For this reason, to simplify the simulations of the $N$-ion model described in Methods, we can make the approximation that the non-radiative decay rate is $\gamma_{\rm nr}\sim 10$ Hz for all ions of this sample.

\onecolumngrid

\begin{figure}[H]
\centerline{\scalebox{0.4}{\includegraphics{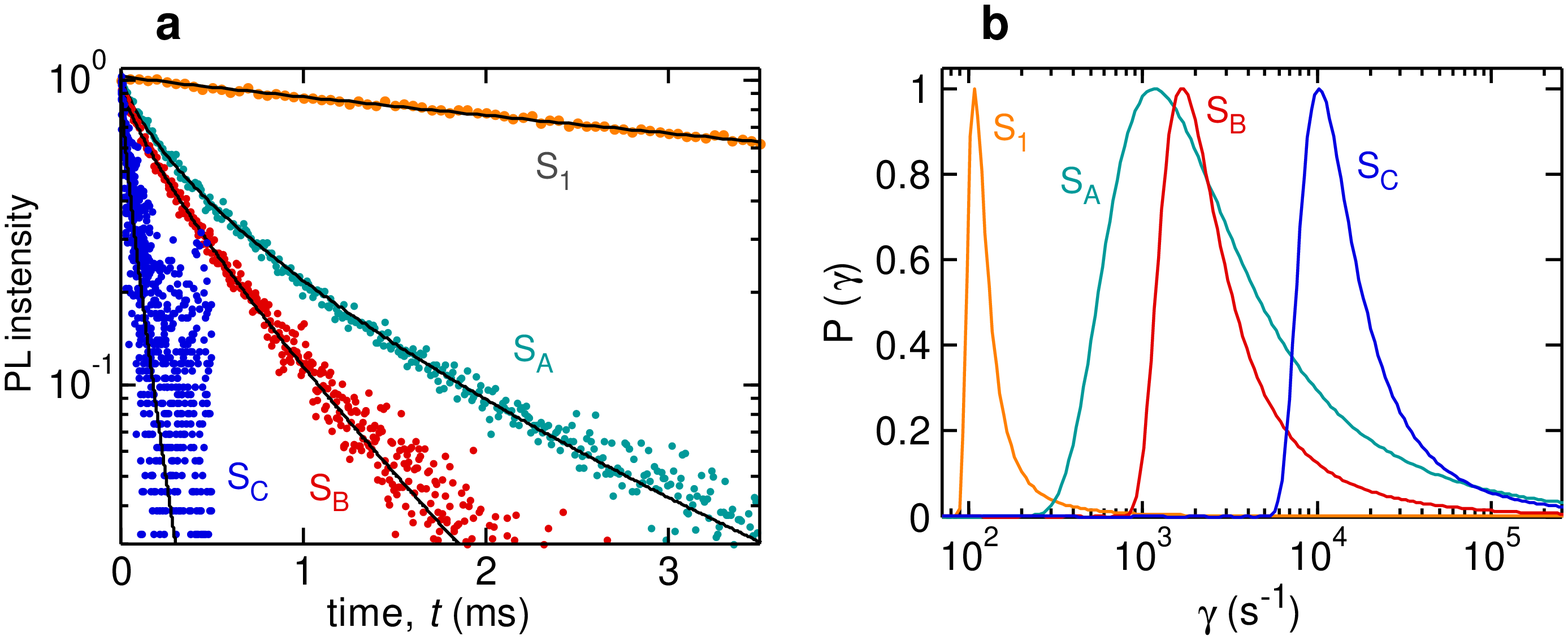}}}
\caption{\textbf{Effect of optimization post-treatment on decay curves.} Comparison between the thin film sample S$_1$ (used for the device of the main text) and three thin film samples made without optimal annealing post-treatment. \textbf{(a)} Measured decay curves. The black solid lines are the best-fit stretched exponential functions. \textbf{(b)} Decay-rate distributions, $P(\gamma)$, obtained by inverse Laplace transformation of the experimental decay curves (see Methods). The distributions are normalized to their respective maxima for a clearer comparison.}
\label{fig:S1_SA_SB_SC}
\end{figure}

\begin{table}[H]
\centerline{\scalebox{0.38}{\includegraphics{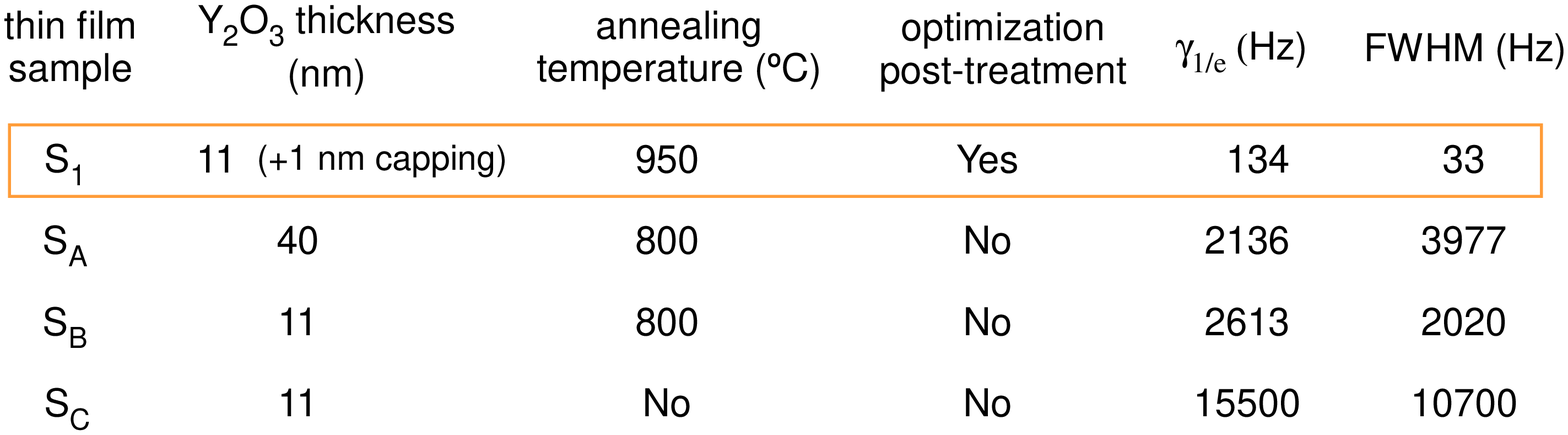}}}
\caption{\textbf{Effect of optimization post-treatment on the decay rates.} Fabrication parameters and fitting parameters of the four thin film samples, where $\gamma_{\rm 1/e}$ is the rate at which de emission decays to 1/e $\sim 0.368$, and FWHM is the full width half maximum of $P(\gamma)$. The sample S$_1$, in which FWHM is only 33 Hz, has an almost mono-exponential decay. Only S$_1$ has a capping layer.}
\end{table}

\vspace{1cm}
\twocolumngrid

\section*{Supplementary Note 2. Characterization of the erbium-graphene hybrid devices.}

We used a scanning confocal microscope setup\cite{Tie:15} to create emission maps of the devices (see Supplementary Fig. \ref{fig:EmissionMap}). This was done by scanning a CW excitation laser at 532 nm over the device and measuring the emitted light at 1545 nm. The area of the graphene monolayer is easily distinguished by the emission quenching effect caused by graphene. We confirmed the graphene monolayer with the Raman spectrum, shown in Supplementary Fig. \ref{fig:Raman}a.

The emission contrast between the regions with and without graphene was used together with the measured decay curves to extract the distribution of decay-enhancement factors, $P(F_{\rm P})$, as described in Methods. The numerical procedure to extract $P(F_{\rm P})$ is valid only if the relationship between erbium emission and excitation laser power is linear. This occurs when the excitation laser power, $P_{\rm exc}$, is sufficiently low to avoid saturation of the the erbium transitions. To verify this, we measured the photon emission as a function of $P_{\rm exc}$, as shown in Supplementary Fig. \ref{fig:Raman}b. We can see that for the laser power used in our experiments, $P_{\rm exc}=0.2$ mW, the relationship between emission and $P_{\rm exc}$ is approximately linear, as required. The experimental data is described by the function $(P_{\rm exc}/P_{\rm sat})/\left[1+(P_{\rm exc}/P_{\rm sat})\right]$ (see Supplementary Reference 4), thus obtaining the saturation laser power $P_{\rm sat}=0.9$ mW (we ignore the dependence of $P_{\rm sat}$ on the excitation laser spot size because this is the same in all our experiments: $\lesssim 1 \mu$m).

Having a low laser power is also important to rule out the possibility of collective erbium-graphene interactions. The erbium concentration is 2\% ($\simeq 10^{21}$cm$^{-3}$), which means that approximately one hundred ions are contained in the plasmon mode volume ($\sim \lambda_{\rm pl}^3$). Using our $N$-ion model (see Methods) and the value of $P_{\rm exc}/P_{\rm sat}$ of our experiments, we have estimated that the average erbium population in the excited state is $\sim$ 0.2\% in the samples with graphene. This implies that there is less than one excited ion within the plasmon mode volume, thus the erbium-graphene interactions can be described by single-ion physics.

\begin{figure}
\centerline{\scalebox{0.35}{\includegraphics{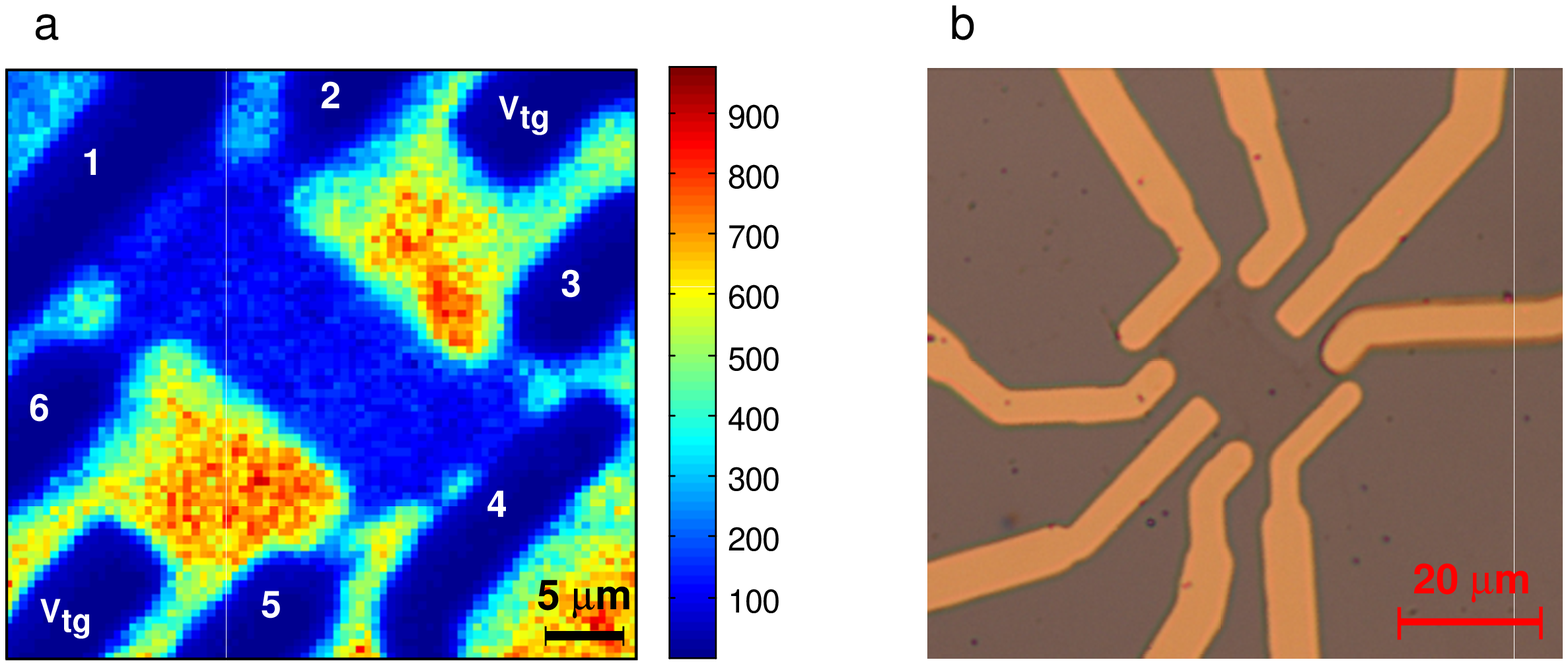}}}
\caption{\textbf{Central area of the erbium-graphene device. }\textbf{(a)} Emission map of the device used in the main text. The central part of the device is a monolayer of CVD-grown, wet-transferred graphene on the erbium-doped thin film S$_1$. Graphene is patterned into a Hall-bar geometry by means of photolithography. The device contains six Cr/Au electrical contacts to graphene (numbered 1 to 6) and two Cr/Au electrical contacts to apply the topgate voltage $V_{\rm tg}$ to the polymer electrolyte. The color scale indicates emission in counts per second. Clearly, emission is quenched by the presence of graphene, which is the result of near-field energy transfer from excited erbium ions to graphene. \textbf{(b)} Optical microcope image of a device. }
\label{fig:EmissionMap}
\end{figure}

\begin{figure}
\centerline{\scalebox{0.29}{\includegraphics{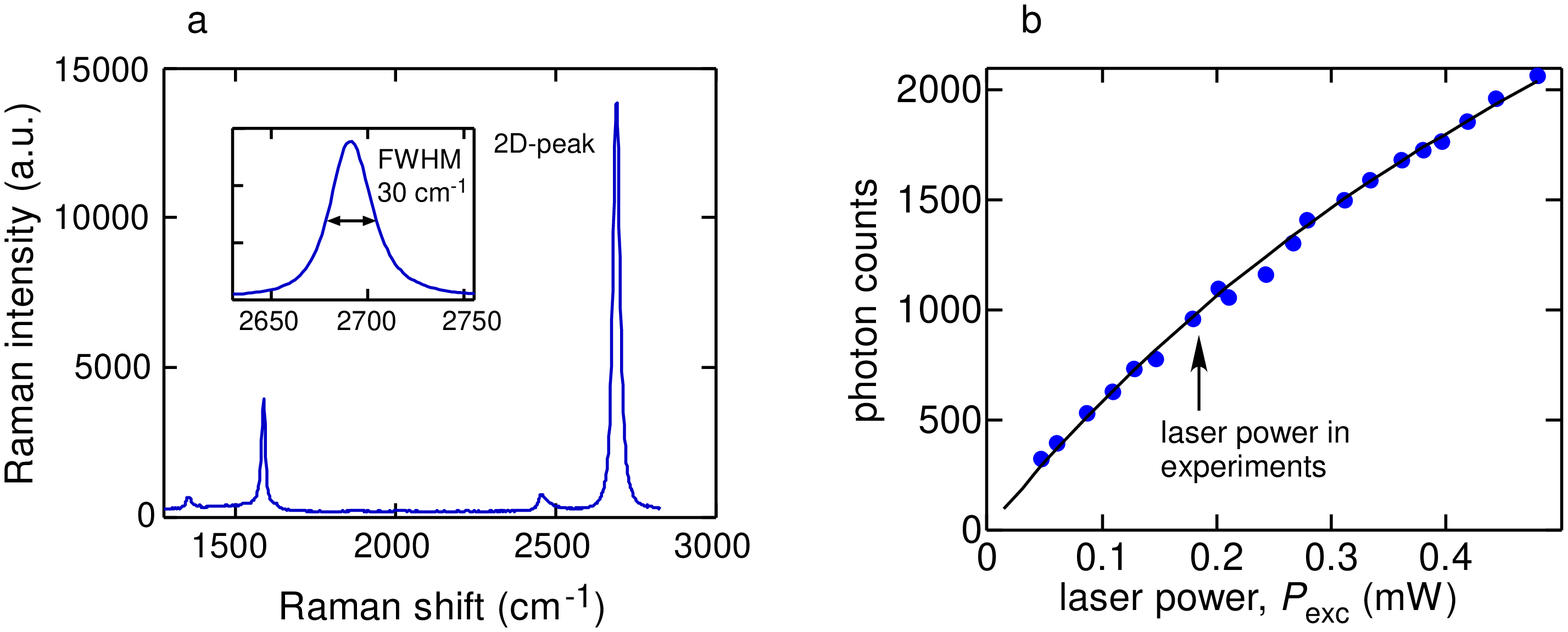}}}
\caption{\textbf{Preliminary tests of the sample.} \textbf{(a)} Graphene characterization using Raman spectroscopy. The FWHM of the 2D peak is approximately 30 cm$^{-1}$, which confirms the presence of the single layer graphene. \textbf{(b)} Photon counts as a function of excitation laser power in the sample S$_1$ without graphene. The relationship between emission and $P_{\rm exc}$ is approximately linear around 0.2 mW, as required for the correct interpretation of emission measurements.}
\label{fig:Raman}
\end{figure}

\section*{Supplementary Note 3. Relationship between the Fermi energy and the applied gate voltages}

It is well established that the two different near-field emitter-graphene coupling regimes occur at different Fermi energies in graphene: below $\sim 0.4$ eV, coupling occurs through interband transitions in graphene, whereas above $\sim 0.6$ eV, coupling occurs by intraband transitions, leading predominantly to plasmons. Between 0.4 eV and 0.6 eV, there is an intermediate regime in which both intraband and interband processes are reduced, which leads to the maximum of emission of far field photons (see Fig. 2a of the main text). In order to confirm that we can statically and dynamically modulate the near-field emitter-graphene coupling between the interband and intraband regimes, we verified the Fermi energy induced by the applied gate voltages for both top and back gates. For this task, we performed a series of independent measurements, including Hall measurements and device resistance measurements.

We performed Hall measurements in order to determine the relationship between the Fermi energy, $E_{\rm F}$, and the topgate voltage, $V_{\rm tg}$. These Hall measurements were carried out in a separate setup, where we applied a current of $I_{\rm bias} = 1 ~\mu$A between the contacts 1 and 4 and measured the Hall voltage between contacts 2 and 6 (see Supplementary Fig. \ref{fig:EmissionMap}), under application of a magnetic field normal to the device surface. We measured the Hall voltage vs. topgate voltage for a positive and a negative field of $B_{+/-} = \pm 0.85$ T, obtaining transverse voltages $V_+$ and $V_-$, respectively. We then obtained the graphene carrier density using $n = I_{\rm bias} (B_+ - B_-)/e(V_+ - V_-)$. By taking the difference for positive and negative perpendicular magnetic field, we removed possible spurious transverse voltages from sample asymmetry. The Fermi energy was obtained from $E_{\rm F} = \hbar v_{\rm F} \sqrt{\pi n}$, where $v_{\rm F}$ is the Fermi velocity. We found that the Fermi energy generated by the topgate can be approximated by $E_{\rm F} \sim E_{\rm F,0} + A V_{\rm tg}$ , with $A = 0.2$ eV V$^{-1}$. This linear relation is valid above $E_{\rm F} \sim 0.3$ eV (see Supplementary Fig. \ref{fig:Calibracion}a). The Fermi energy at zero gate voltage $E_{\rm F,0}$ varies significantly from device to device as it is determined by the amount of charge impurities (see Supplementary Fig. \ref{fig:Calibracion}b). Therefore, the slope $A$ is the most important parameter we extracted from the Hall measurements. In agreement with previous studies, we find that the Fermi energy induced by the polymer electrolyte gate cannot be approximated by a straightforward capacitive coupling constant (which would give a carrier density that is linear in gate voltage rather than a Fermi energy linear in gate voltage).

For the dynamic control of the erbium-graphene interactions, we need to know the Fermi energy oscillation amplitude, $\Delta E_{\rm F}$, that is induced by the AC voltage amplitude applied to the backgate, $\Delta V_{\rm bg}$. Just before every dynamic modulation experiment, we found the relationship between $\Delta E_{\rm F}$ and $\Delta V_{\rm bg}$ by means of a procedure based on device resistance measurements. To illustrate our procedure, we show in Supplementary Figs. \ref{fig:Calibracion}c-d the device resistance measurements that we did for the dynamic modulation experiments of Fig. 3 of the main text. We first measured the device resistance as a function of $ V_{\rm tg}$ while keeping the backgate voltage at 0 V, as shown in Supplementary Fig. \ref{fig:Calibracion}c. Then, we converted $V_{\rm tg}$ to $E_{\rm F}$ by using the calibrations described in the previous paragraph, thus obtaining the device resistance versus $E_{\rm F}$. Next, we tuned $V_{\rm tg}$ to the base Fermi energy of the corresponding dynamic modulation experiment. This was $V_{\rm tg} = 0.2$ V ($E_{\rm F} = 0.45$ eV) in the measurements of Fig. 3 of the main text, and $V_{\rm tg} = 2.2$ V ($E_{\rm F} = 0.8$ eV) in the measurements of Fig. 4 of the main text. Then, while keeping $V_{\rm tg}$ fixed, we measured the device resistance versus $V_{\rm bg}$, and converted the device resistance into Fermi energy using the measurement of Supplementary Fig. \ref{fig:Calibracion}c, thus finding $\Delta E_{\rm F}$ versus $\Delta V_{\rm bg}$.

\begin{figure}
\centerline{\scalebox{0.33}{\includegraphics{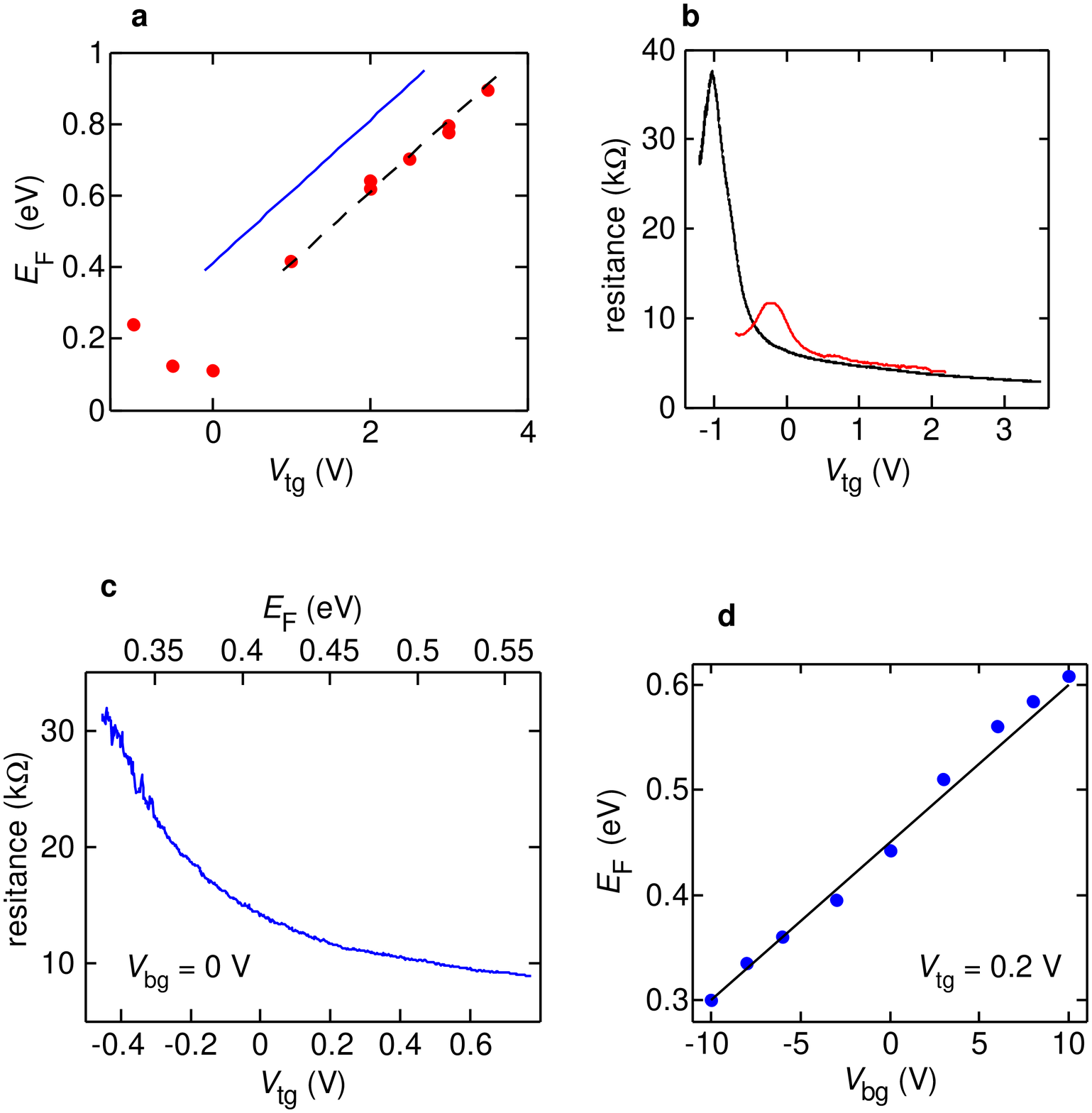}}}
\caption{\textbf{Calibration of the Fermi energy of graphene.} \textbf{(a)} Calibration of the Fermi energy as a function of the topgate voltage. The calibration is based on the Fermi energy versus topgate voltage obtained experimentally by means of Hall effect measurements (red dots). These measurements are described by a linear function, $E_{\rm F} \sim E_{\rm F,0} + A V_{\rm tg}$, where $A = 0.2$ eV V$^{-1}$ (black dashed line). The Fermi energy at zero gate voltage $E_{\rm F,0}$ varies significantly from device to device. In the case of the device used in the main text, $E_{\rm F,0} = -0.41$ eV, and the corresponding linear function is plotted as a blue solid line. \textbf{(b)} Device resistance as a function of topgate voltage for two different devices, from which we can determine the Fermi Dirac point of the device. \textbf{(c)} Device resistance as a function of $E_{\rm F}$ just before modulation measurements of Figure 3 of the main text. The curve was obtained by measuring the device resistance as a function of $V_{\rm tg}$ for $V_{\rm bg} = 0$ V, and converting $V_{\rm tg}$ to $E_{\rm F}$ using the calibration from (a). \textbf{(d)} Fermi energy versus $V_{\rm bg}$, with fixed topgate, $V_{\rm tg} = 0.2$ V ($E_{\rm F} = 0.45$ eV). Here, the experimental points were obtained by measuring the device resistance for different backgate voltages, and converting device resistance into $E_{\rm F}$ using the curve of (c). The linear fit (black solid curve) has a slope of $B = 15$ mV eV$^{-1}$, which we use to know the Fermi energy in the experiments of Figure 3 of the main text.}
\label{fig:Calibracion}
\end{figure}

By means of numerical fitting, we found $\Delta E_{\rm F} \sim B \Delta V_{\rm bg}$ , with $B = 15$ eV mV$^{-1}$ (13 eV mV$^{-1}$) for a topgate-generated Fermi energy of 0.45 eV (0.8 eV), which corresponds to a capacitance per unit area of $C = 1.6 \cdot 10^{-3}$ Fm$^{-2}$  ($2.4 \cdot 10^{-3}$ Fm$^{-2}$). This capacitance has been calculated using the expression $B = (\pi C \hbar^2 v_{\rm F}^2) / (2 e E_{\rm F} )$, where $E_{\rm F}$ is the base Fermi energy induced by the topgate voltage, and $e$ is the electron charge. This expression comes from the first-order term of the Taylor expansion of the Fermi energy as a function of gate voltage\cite{Cas:09}. It is remarkable that the experimental backgate capacitance of our devices is much larger than that expected for the 285-nm-thick SiO$_2$ layer of our substrate: $C_{\rm SiO_2} = 1.2 \cdot 10^{-4}$ Fm$^{-2}$. This indicates that the electrolyte in contact with graphene has an enhancement effect on the backgate capacitance. A similar enhancement effect has been observed in previous works \cite{Xia:10, Gro:17}, and is likely related to the large effective epsilon induced by the presence of the ions of the polymer electrolyte directly above the graphene layer.

We carried out dynamic modulation experiments for different backgate-voltage amplitudes, as shown in Supplementary Fig. \ref{fig:Modulation_BackGateAmplitude}, to confirm that we can modulate the Fermi energy between the interband and the intraband regimes. In these measurements, $\Delta V_{\rm bg}$ was varied between 3 V and 10 V, while the topgate voltage was fixed at $V_{\rm tg} = 0.2$ V ($E_{\rm F} = 0.45$ eV). We theoretically simulated the emission by using our $N$-ion model (see Methods) with the ion density distribution of Fig. 2c of the main text. In each simulation, $\Delta E_{\rm F}$ was a free parameter. Supplementary Figure \ref{fig:Modulation_BackGateAmplitude}e shows the best-fit values of $\Delta E_{\rm F}$ for the different applied voltage amplitudes, $\Delta V_{\rm bg}$. The relationship between $\Delta E_{\rm F}$ and $\Delta V_{\rm bg}$ is approximately linear, with a slope of $B = 15$ eV mV$^{-1}$, the same value as obtained from the device resistance measurements described in the previous paragraph. This corroborates the calibration with the device resistance measurements.

We performed a dynamic modulation test to verify the transition from the interband regime to the intraband regime as the Fermi energy is increased. The test consisted in a series of measurements, shown in Supplementary Fig. \ref{fig:Modulation_TopGate}, in which the base Fermi energy was increased from 0.4 eV to 0.7 eV using the topgate. A small modulation amplitude of $\Delta E_{\rm F} \sim \pm 60$ meV was applied using the backgate. We can see that, as the base Fermi energy increases, the oscillation inverts its sign, thus indicating the transition into the intraband regime, in which plasmon creation is the dominant decay mechanism of the erbium ions. This sign inversion corresponds to the change of the sign of the slope of the emission contrast shown in Fig. 2a of the main text.

We checked the Fermi energy during optical measurements by monitoring the device resistance (see Supplementary Fig. \ref{fig:dummy}). We observed that the device resistance remains sufficiently stable for several hours, which indicates a high degree of stability of the Fermi energy induced by the applied gate voltages. This long-term stability allowed the realization of optical measurements during several hours without significant Fermi energy drifts.

\begin{figure}[H]
\centerline{\scalebox{0.37}{\includegraphics{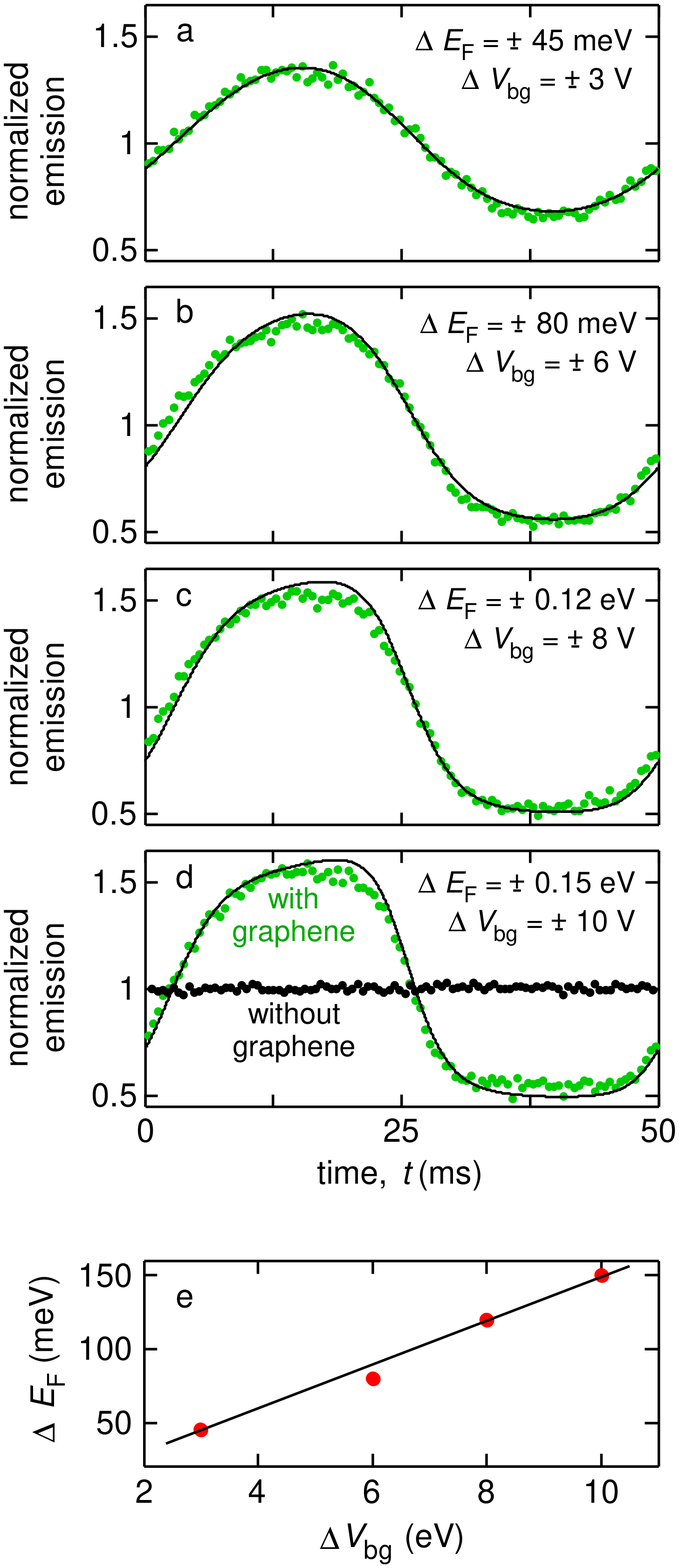}}}
\caption{\textbf{Verification of the modulation between the interband and intraband regimes.} \textbf{(a-d)} Dynamic modulation measurements (green dots) for different backgate modulation amplitudes, $\Delta V_{\rm bg}$, between $\pm$3 V and $\pm$10 V. In all measurements the topgate voltage is 0.2 eV, which corresponds to a base Fermi energy of $E_{\rm F} = 0.45$ eV. The solid black lines show the theoretical emission simulated using the $N$-ion model with $\Delta E_{\rm F} \sim B V_{\rm bg}$ , where $B = 15$ eV mV$^{-1}$. \textbf{(d)} The emission measurement on graphene (green dots) is compared with the emission measurements when the excitation laser shines a region of the device without graphene (black dots). We use this comparison to confirm that emission oscillations are caused by graphene and not by any other effect induced by the gates. \textbf{(e)} Best-fit values of $\Delta E_{\rm F}$ obtained using the $N$-ion model with the oscillations of subfigures a-d.}
\label{fig:Modulation_BackGateAmplitude}
\end{figure}

\begin{figure}[H]
\centerline{\scalebox{0.37}{\includegraphics{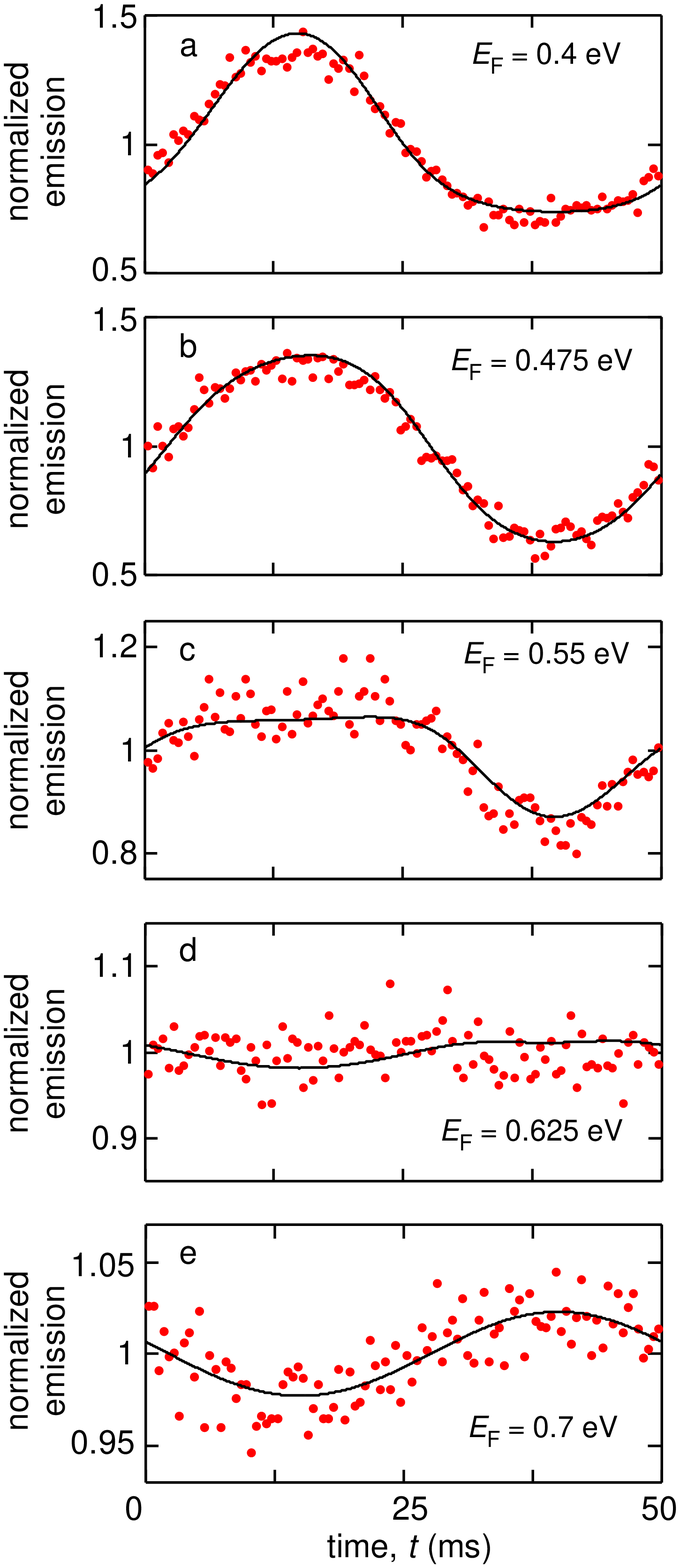}}}
\caption{\textbf{Inversion of the modulation signal due to plasmons.} Dynamic modulation measurements (red dots) for a Fermi energy modulation amplitude of $\Delta E_{\rm F} = \pm 60$ meV, induced by an AC backgate voltage at 20 Hz. The base Fermi energy induced by the topgate is \textbf{(a)} 0.4 eV, \textbf{(b)} 0.475 eV, \textbf{(c)} 0.55 eV, \textbf{(d)} 0.625 eV and \textbf{(e)} 0.7 eV. The sign of the oscillation inverts as the base Fermi energy increases, which is an obvious signature of the transition into the intraband regime in graphene. The black solid curves show the numerical simulation using the 50-ion model described in Methods.}
\label{fig:Modulation_TopGate}
\end{figure}

\begin{figure}[H]
\centerline{\scalebox{0.23}{\includegraphics{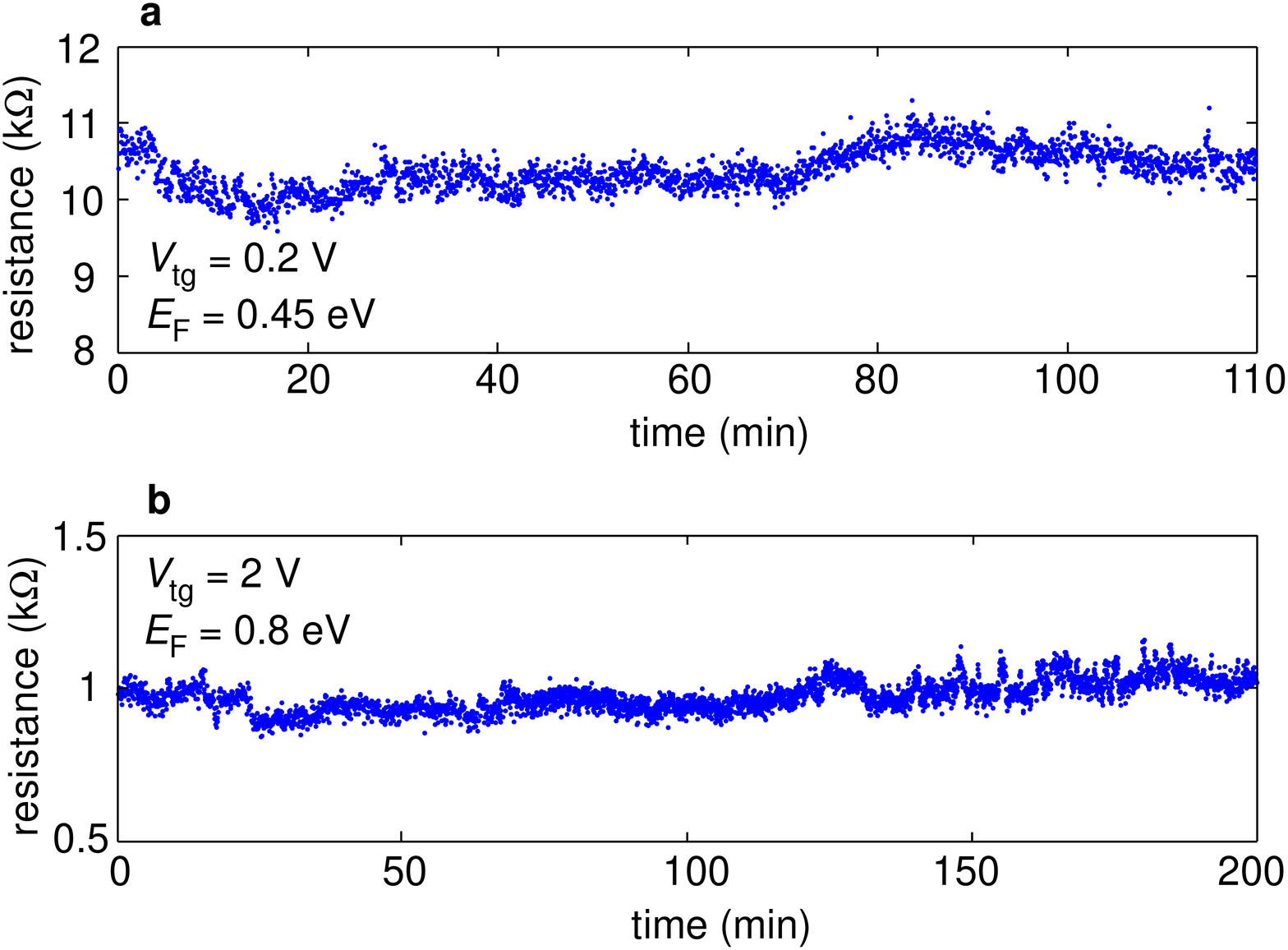}}}
\caption{\textbf{Monitoring the stability of the Fermi energy.} Measurements of the device resistance during optical measurements at Fermi energies of \textbf{(a)} 0.45 eV and \textbf{(b)} 0.8 eV. We applied a potential difference over the graphene sheet (typically, 2 mV) in order to monitor the device resistance and verify, in this way, that the Fermi energy remains stable during the measurements.} \label{fig:dummy}
\end{figure}

\vspace{4mm}

\section*{Supplementary Note 4. Results obtained with other erbium-graphene devices.}

\begin{figure}[H]
\centerline{\scalebox{0.34}{\includegraphics{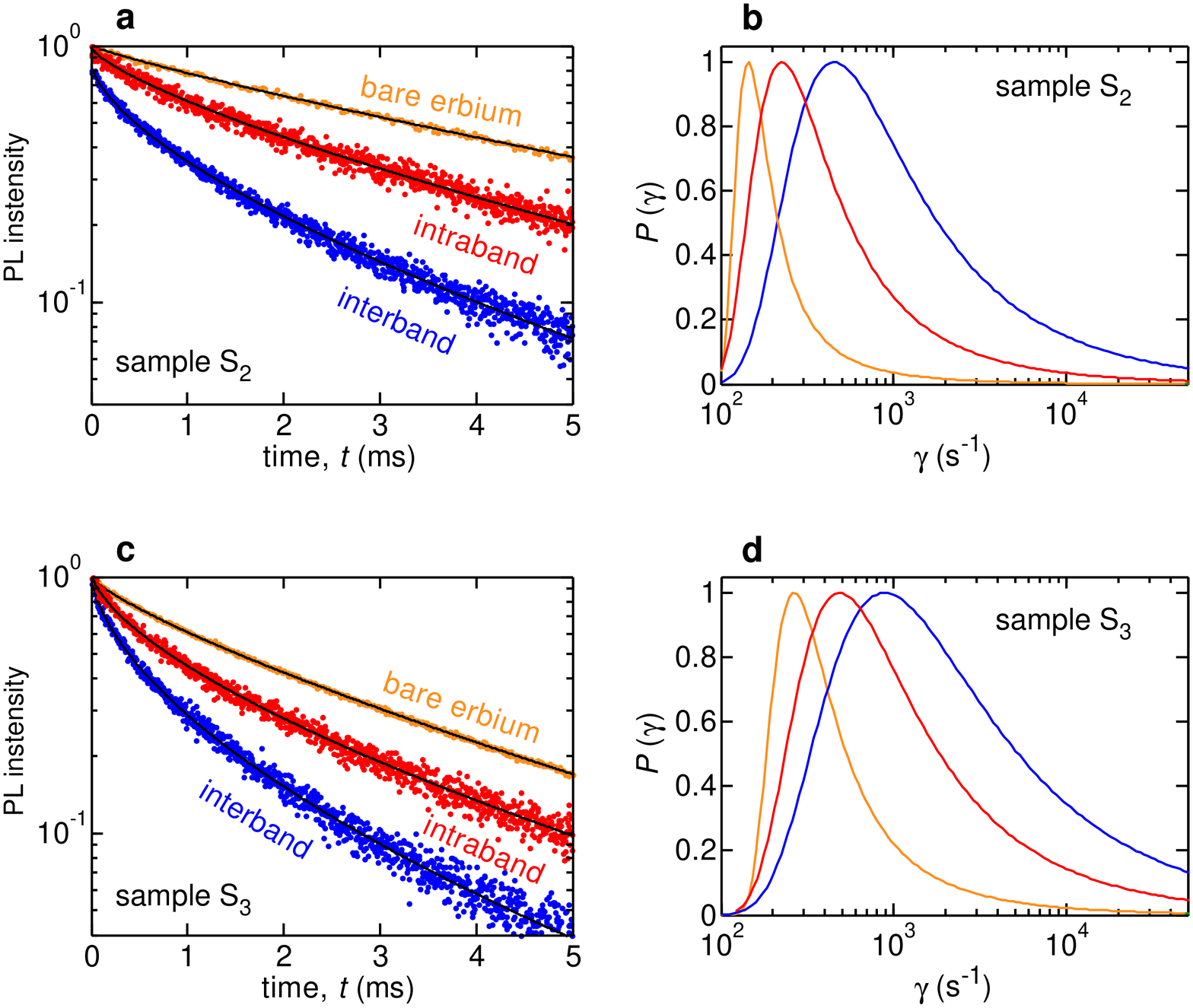}}}
\caption{\textbf{Optical characterization of additional erbium-graphene devices.} Measured decay curves and their corresponding decay rate distributions for two devices on two different thin film samples, that we denote as S$_2$ and S$_3$.  \textbf{(a-b)} Results for the device on S$_2$, which was fabricated in the same way as S$_1$, but without the capping layer. \textbf{(c-d)} Results for the device on S$_3$, which was fabricated by growing a 11-nm-thick Y$_2$O$_3$:Er (2\%) film on a Si(100) substrate and subsequent annealing at 1000~$^\circ$C. For each device, we considered three cases: graphene in the interband regime ($E_{\rm F} = 0.2$ eV, blue), graphene in the intraband regime ($E_{\rm F} = 0.8$ eV, red), and without graphene (orange). The black solid lines correspond to the best-fit stretched-exponential curves. The decay rate distributions were obtained by inverse Laplace transformation of the measured decay curves (see Methods).}
\label{fig:S2_S3}
\end{figure}

We obtained decay curves and emission maps from fifteen graphene devices on six different thin film samples. Furthermore, we performed dynamic modulation in three different devices on two different thin film samples (the first thin film samples, which were grown on Si(100), did not have a backgate). We measured similar erbium-graphene interactions in all devices. All of them showed a clear transition between the interband and intraband regimes. Supplementary Figure \ref{fig:S2_S3} shows the decay curves and the corresponding decay rate distributions of two devices.

\vspace{4mm}

\section*{Supplementary Note 5. Maximum decay-enhancement factor that can be extracted from the decay curves.}

The decay-enhancement-factor distributions, $P(F_{\rm P})$, and the density distribution, $P(z)$, were obtained by analyzing the experimental decay curves and the emission contrast measurements together, as described in Methods. The decay curves, $n(t)$, reflect mainly the dynamics of the ions with the lowest decay enhancements factors since these are the ions that emit the largest amount of photons during lifetime measurements. Using the numerical techniques of Supplementary Reference 8, we determined the maximum decay-enhancement factor that can be extracted from the measured decay curves, $F_{\rm P, max} \sim 1,000$, which corresponds to an erbium-graphene separation of $\sim$7 nm (see Supplementary Fig. \ref{fig:cutoff}a). To check that this value of $F_{\rm P, max}$ actually represents the accuracy limit provided by our experimental decay curves, we did a test which consisted in calculating the decay curves from $P(F_{\rm P})$ by doing the inverse numerical procedure that we did to calculate $P(F_{\rm P})$ from the experimental decay curves and emission contrast measurements (see Methods). In this test, we first converted $P(F_{\rm P})$ into $P(\gamma)$, and then we calculated the decay curves by numerically computing the Laplace transformation,
\begin{equation}
n(t)=\int_{0}^{\gamma_{\rm max}} \frac{P(\gamma)}{\gamma} \textrm{e}^{-\gamma t} d\gamma  , \nonumber
\label{Eq_Laplace}
\end{equation} \noindent
where $\gamma_{\rm max}$ is the maximum decay rate considered in the decay curve. For this test, we neglected the intrinsic non radiative decay of the erbium-doped thin film. Supplementary Figure \ref{fig:cutoff}b shows the beginning of the decay curve of the device on the thin film S$_1$ in the interband regime, together with the decay curves calculated from $P(\gamma)$ for different $\gamma_{\rm max}$. As $\gamma_{\rm max}$ is increased, the calculated decay curves get closer to the experimental decay curves. We can see that the ions with $F_{\rm P} > 1,000$ have a very small effect which appears only in the first microseconds of the decay curves. To experimentally observe the contribution of these ions more clearly, we would need to reduce the histogram bin size as well as to increase the time of the measurements. However, given the very low photon emission of our devices, that would require weeks of continuous photon collection. It is more convenient to determine the distributions of the ions with $F_{\rm P} > 1,000$ by using the emission contrast measurements, as described in Methods. The maximum cutoff factor $F_{\rm P, max} \sim 1,000$ corresponds to an erbium-graphene separation between 5 nm and 8 nm, depending on the Fermi energy (see Supplementary Fig. \ref{fig:cutoff}a). As a trade-off, for our model we use 7 nm as the distance below which the density distribution is extracted from the emission contrast measurements. In this way, we obtain the erbium-graphene distances $z_i$ ($i$ = 1,...,50) shown in Supplementary Fig. \ref{fig:cutoff}c. The distances $z_i$ are converted into the the density distribution $P(z)$ and vice versa by means of numerical integration and discretization, respectively.

It is interesting to remark the high $F_P$ factors at short distances from graphene. As we can see in Supplementary Fig. \ref{fig:cutoff}a, $F_{\rm P}$ is higher than $10^6$ at separations below 2 nm. This corresponds to energy transfer rates higher than $\gamma_{\rm gr} \simeq F_{\rm P} \gamma_{\rm ed} \sim 75$ MHz, where $\gamma_{\rm ed}\sim$75 Hz for erbium ions\cite{Web:68}. Nonetheless, these values do not represent the ultimate limit of the emitter-graphene interactions. In fact, $\gamma_{\rm gr}$ can be further increased by more than one order of magnitude by patterning the graphene monolayer into waveguides and cavities with defined plasmonic modes \cite{Kop:11,Gon:16,Chr:12} or by using a graphene-insulator-metal heterostructure\cite{Alc:18}. Such high values of $\gamma_{\rm gr}$ can be modulated with graphene, in which $f_{\rm mod}$ can be up to tens of GHz\cite{Pha:15} (In general, $\gamma_{\rm gr}$ determines the maximum modulation frequency that the ion dynamics can follow). The experimental observation of such high modulation frequencies may involve thinner erbium thin films and a configuration based on graphene plasmon cavities with optical nanoantennas or waveguides\cite{Tiecke:15} that enhance the far-field emission of the ions with very large $\gamma_{\rm gr}$.

\onecolumngrid

\begin{figure}[H]
\centerline{\scalebox{0.39}{\includegraphics{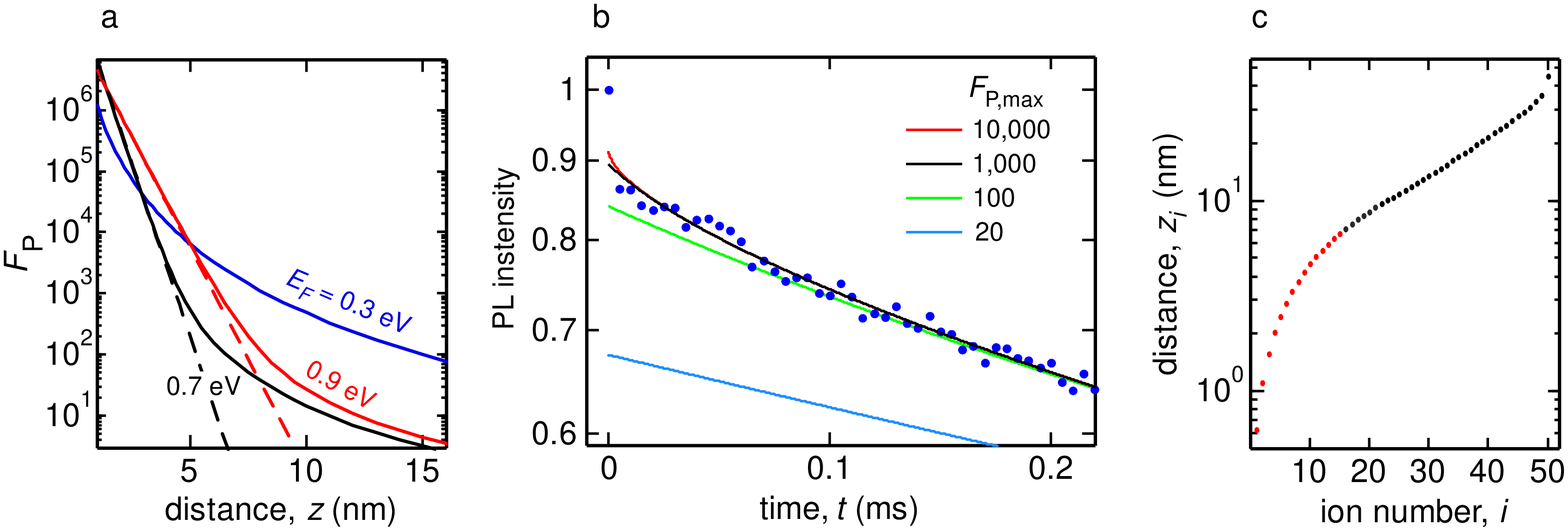}}}
\caption{\textbf{Evaluation of the maximum decay-enhancement factor.} \textbf{(a)} Decay-enhancement factor $F_{\rm P}(z)$ as a function of erbium-graphene distance for three different Fermi energies: $E_{\rm F} = $0.3 eV (blue), 0.7 eV (black) and 0.9 eV (red). These theoretical functions have been calculated as in Supplementary Refs. 9 and 10, using the experimental parameters described in Methods. The dashed lines represent the ideal exponential decay, $F_{\rm P} \propto \exp(-4 \pi z / \lambda_{\rm pl})$, of the long-distance propagating plasmons without losses, where $\lambda_{\rm pl}= 4.9$ nm (7.7 nm) for $E_{\rm F} = 0.7$ eV (0.9 eV). The deviations from the ideal plasmon exponential decay are caused by typical losses in CVD graphene. \textbf{(b)} Evaluation of the maximum decay-enhancement factor provided by the decay curves. The blue dots show the beginning of the measured decay curve of the device on thin film sample S$_1$ in the interband regime (see Fig. 2a of the main text). The solid lines are the theoretical decay curves calculated from $P(F_{\rm P})$ considering only the ions with $F_{\rm P}$ below a maximum cutoff factor, $F_{\rm P, max}$. \textbf{(c)} Erbium-graphene distances used to simulate the emission during dynamic modulation of the near field. The distances $z_i$ are obtained either from the decay curves (black dots) or from the emission contrast (red dots).}
\label{fig:cutoff}
\end{figure}

\twocolumngrid

\section*{Supplementary Note 6. Time-of-Flight Secondary Ion Mass Spectrometry (ToF-SIMS)}

\vspace{-2mm}

We obtained the density profile of the Er$^{3+}$ ions by ToF-SIMS (see Methods). We determined the Er$^{3+}$ density profile from the removed YO$^-$ particles instead of from the ErO$^-$ particles because the YO$^-$ signal was much less noisy. The noise of the ErO$^-$ signal was relatively high because we had to use the less abundant isotope $^{168}$Er, with 27\% abundance, in order to prevent overlap with other species. To verify that the YO$^-$ signal provides a good estimation of the Er$^{3+}$ density profile we plot in Supplementary Fig. \ref{fig:TOF_SIMS} the density profiles obtained from both ErO$^-$ and YO$^-$ signals. In this figure we can see that the diffusion is practically the same for both species.

\vspace{-2mm}

\begin{figure}[H]
\centerline{\scalebox{0.36}{\includegraphics{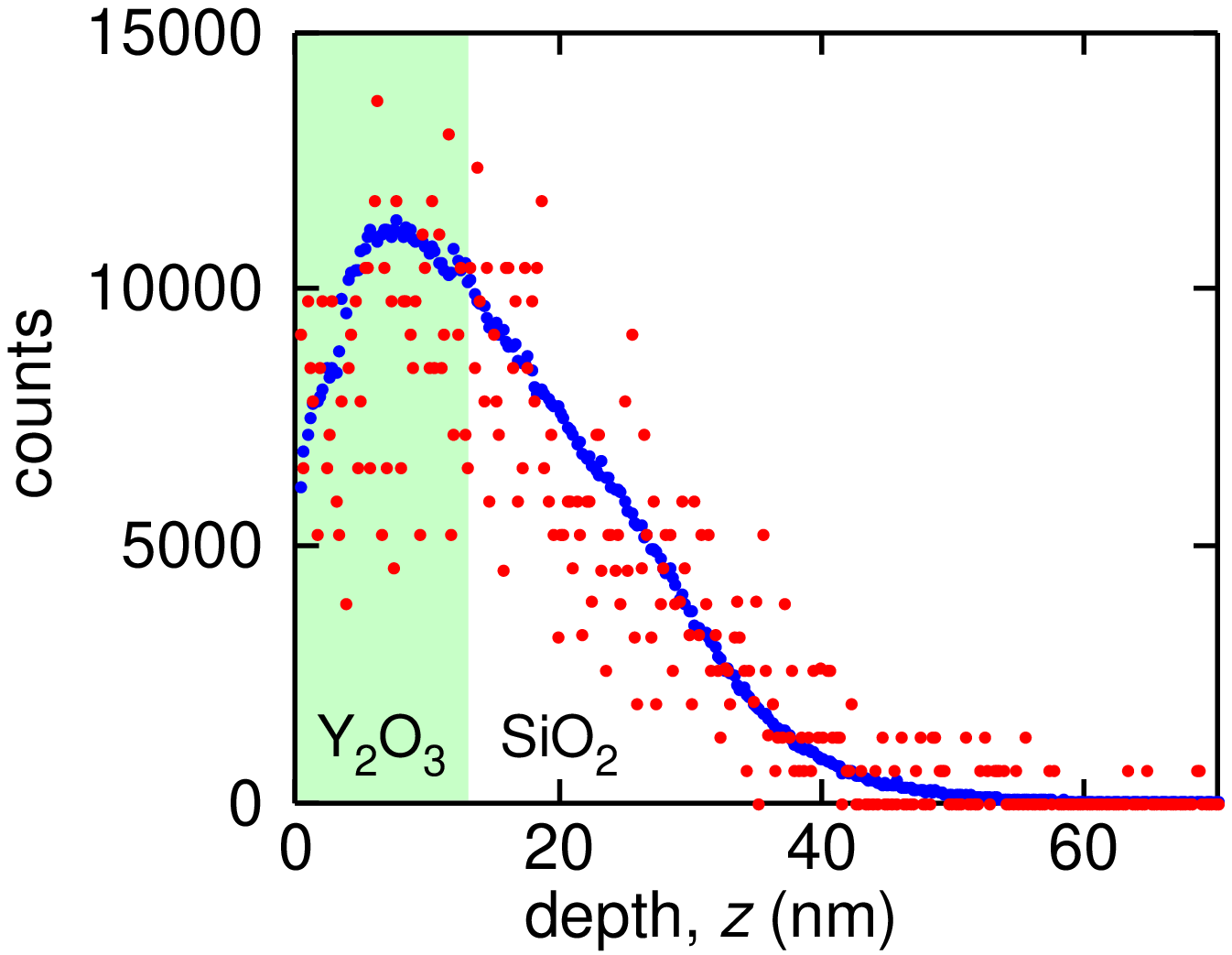}}}
\caption{\textbf{TOF-SIMS measurements.} Depth profiles of YO$^-$ (blue dots) and ErO$^-$ particles (red dots). The similarity between both profiles proves that the diffusion is very similar for both Y$^{3+}$ and Er$^{3+}$ ions. The Erbium signal has been multiplied by 650 for a clear comparison.}
\label{fig:TOF_SIMS}
\end{figure}
